\documentclass[prl,showpacs,showkeys,twocolumn,notitlepage,10pt]{revtex4-1}
\usepackage{amsfonts,bbold,amsmath,amssymb,graphicx,epstopdf,verbatim,dsfont,color}
\usepackage[english]{babel}
\newcommand{\tr}{{\rm Tr}}

\newcommand{\be}{\begin{equation}}
\newcommand{\ee}{\end{equation}}
\newcommand{\beq}{\begin{eqnarray}}
\newcommand{\eeq}{\end{eqnarray}}

\begin{document}
\title{Minimizing irreversible losses in quantum systems by local counter-diabatic driving }
\author{Dries Sels}
\affiliation{Department of Physics, Boston University, Boston, MA 02215, USA}
\affiliation{TQC, Universiteit Antwerpen, B-2610 Antwerpen, Belgium}
\author{Anatoli Polkovnikov}
\affiliation{Department of Physics, Boston University, Boston, MA 02215, USA}

\date{\today}

\maketitle

{\bf Counter-diabatic driving protocols were proposed as a means to do fast changes in the Hamiltonian without exciting transitions~\cite{demirplak_03, demirplak_05, berry_09}. Such driving in principle allows one to realize arbitrarily fast annealing protocols or implement fast dissipationless driving, circumventing standard adiabatic limitations requiring infinitesimally slow rates. These ideas were tested and used both experimentally and theoretically in small systems, but in larger chaotic systems it is known that exact counter-diabatic protocols do not exist. In this work we develop a simple variational approach allowing one to find best possible counter-diabatic protocols given physical constraints like locality. These protocols are easy to derive and implement both experimentally and numerically. We show that, using these approximate protocols, one can drastically decrease dissipation and increase fidelity of quantum annealing protocols in complex many-particle systems. In the fast limit these protocols provide an effective dual description of adiabatic dynamics where the coupling constant plays the role of time and the counter-diabatic term plays the role of the Hamiltonian.
}

\section{Introduction}
Despite the time-reversal symmetry of the microscopic dynamics of isolated systems, losses are ubiquitous in any process that tries to manipulate them. Whether it's the heat produced in a car engine or the decoherence of a qubit, all losses arise from our lack of control on the microscopic degrees of freedom of the system. Since the early-days of thermodynamics, and actually even before, the adiabatic process has emerged as a universal way to minimize losses, leading to the concept of Carnot efficiency -- the cornerstone of modern thermodynamics. In spite of its conceptual importance, practical implications of the Carnot efficiency are limited since the maximal efficiency goes hand in hand with zero power.  Nonetheless, by sacrificing some of the efficiency one can run the same Carnot cycle at finite power (see e.g. Ref.~\cite{curzon_75}). Heat engines might appear a problem of the past but the understanding of finite-time thermodynamics in small (quantum) systems has become increasingly important due to developments in quantum information and nanoengineering. 

Developing and understanding methods to induce quasi adiabatic dynamics at finite times is paramount to the advancement of quantum information technologies. In general, one could distinguish between two (complementary) approaches. On one hand one can, for a fixed setup, try to develop optimal driving protocols that result in minimal loss under certain constraints. Such protocols were recently suggested as appropriate geodesic paths in the parameter space, both in the context of thermodynamics~\cite{sivak_12} and in the context of adiabatic state preparation~\cite{zanardi_09,tomka_16}. The optimal protocols were also analyzed numerically using various optimum control ideas~\cite{rahmani_11, rahmani_13, martinis_14, masuda_14, karzig_15}. On the other hand, one can try to engineer fast non-adiabatic protocols that lead to the same result as the fully adiabatic protocol. In particular, transitionless driving protocols were recently proposed and explored in small single-particle systems~\cite{demirplak_03, demirplak_05, berry_09, deffner_14, campbell_15, baksic_16} with numerical extensions to larger interacting  systems~\cite{saberi_14}. In this approach one introduces an auxiliary counter-diabatic (CD) Hamiltonian drive on top of a target Hamiltonian to suppress all transitions between eigenstates.  A general problem with this approach is that in complex chaotic systems the exact CD Hamiltonian is non-local and exponentially sensitive to any tiny perturbations. The goal of this work is to overcome these difficulties by providing a new route to finding approximate optimal CD driving protocols by restricting to a class of physical operators, e.g. those accessible in experiments. In this work we focus specifically on local CD protocols.

\begin{figure}
  \centering
  \includegraphics[width=\linewidth]{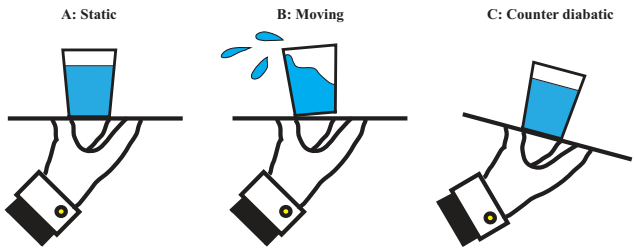}
  \caption{{\bf The counter diabatic waiter.} A waiter's goal is to deliver a tray with a glass of water from the bar to a costumer without spilling. In the beginning and the end of the task the system should look like situation A. An adiabatic waiter can always be in situation A, but with a desire to be more efficient and speed up the protocol, a naive waiter will find himself in the undesirable situation B somewhere during the task. By tilting the tray (situation C), an example of counter-diabatic driving, situation B can be avoided and the desired tasks can be achieved much faster.}
  \label{fig:waiter}
\end{figure}

The ideas of CD driving are certainly not new and are used on everyday basis in nature. Let us illustrate these ideas using an example of a waiter bringing a tray with a glass of water from the bar to a table (see Fig.~\ref{fig:waiter}). As we will show in this work, this simple example contains very important insights, which will become relevant to the core of this paper. The goal of the waiter is to deliver the water to the table with a high fidelity, i.e. without spilling or splashing it. Of course the glass should be vertical in the beginning of the process, i.e. when the waiter is leaving the bar and at the end of the process, when the waiter reaches the table. The simplest protocol which can be adopted by the waiter is adiabatic, where he slowly moves along the shortest path (geodesic) connecting the bar and the table keeping the tray vertically at all times. This will work but will require a lot of time and thus the efficiency of such ``adiabatic waiter'' will be very low. An efficient waiter has to serve more customers by going faster and this requires a different tactic. When accelerating to reach a finite speed, a pseudo-force will act on the drinks, which will cause the drinks to spill or even tip over. This can be avoided by acting on the drinks with an equal and opposite force and that's exactly what waiters do. Moreover, the same tilt can counter a drag force caused by the air if the waiter runs very fast. In fact, by tilting the tray while moving the waiter induces a CD force. Despite the fact that the system of the tray, the glass and water is complex and chaotic it is clear from our everyday experience that this CD protocol can be extremely efficient. Let us highlight several important points, which we can learn from this intuitive example. We will come back to these points later, when we discuss various physical examples:

\begin{itemize}

\item In order to implement an efficient CD protocol one has to introduce new degrees of freedom (like a tilt), which do not show up in the initial and final state as well as in the adiabatic path.

\item The system does not generally follow an instantaneous ground state: at intermediate times the waiter tilts the tray and moves it fast, which corresponds to a highly excited state of the system in the lab frame.

\item The CD protocol corresponds to adding local terms to the Hamiltonian of the system like the gravitational field. This protocol is only sensitive to the velocity and acceleration of the waiter.

\end{itemize}

As we will show these observations underlie crucial ideas behind engineering CD protocols in complex systems such as locality and gauge equivalence. Using these ideas as a guiding principle, we develop a simple variational approach allowing one to find local and robust approximate counter-adiabatic Hamiltonians. These counter-terms allow one to achieve truly spectacular results in suppressing dissipation or targeting ground states of gapped or gapless many-particle systems with a very high fidelity at a very fast speeds. An important advantage of the variational method is that it allows one to find efficient CD protocols without the need of diagonalizing the Hamiltonian, in particular, in the thermodynamic limit. Moreover one can check the accuracy of the variational ansatz by analyzing the stability of the protocol with respect to adding additional terms.

\section{Local counter-diabatic driving.}

\subsection{CD driving in quantum and classical systems}

Let's have a closer look at how transitions between eigenstates actually arise and how one can suppress them. Consider a state $\left| \psi \right\rangle$, evolving under the Hamiltonian $H_0\left(\lambda(t) \right)$, which is time dependent through the parameter $\lambda(t)$. In general, $\lambda$ can be a multicomponent vector parameter (for example in the case of a waiter $\lambda$ can stand for his $x$ and $y$ coordinates), but in this work we will focus on the single-component case to avoid extra complications. If the parameter changes in time then for a moving observer in the instantaneous eigenbasis of $H_0$, the laws of physics are modified. This is of course very well known for the case of an accelerated or a rotating frame, but in fact it applies to all types of motion. Specifically, the Hamiltonian picks up an extra contribution and becomes
\begin{equation}
H^{\rm eff}_0=\tilde H_0-\dot{\lambda} \tilde{\mathcal A}_\lambda,
\end{equation}
here $\tilde{\mathcal A}_\lambda$ is the adiabatic gauge potential in the moving frame. It is geometric in origin and related to the infinitesimal transformations of the instantaneous basis states in the quantum case and to the infinitesimal canonical transformations of conjugate variables (like coordinates and momenta) in the classical case (see methods and Ref.~\cite{kolodrubetz_16} for details).

In the moving frame the Hamiltonian $\tilde H_0$ is diagonal (stationary), so all non-adiabatic effects must be due to the second term.  The idea of the CD driving is to evolve the system with the Hamiltonian 
\[
H_{\rm CD}(t)=H_0+\dot\lambda \mathcal A_\lambda
\]
such that in the moving frame $H_{\rm CD}^{\rm eff}(t)=\tilde H_0$ is stationary and no transitions occur. Note that by construction in the zero velocity limit $|\dot\lambda|\to 0$ the CD Hamiltonian $H_{\rm CD}(t)$ reduces to the original Hamiltonian $H_0(t)$ as expected. It is easy to show that the gauge potential $\mathcal A_\lambda$ satisfies the following equation (see methods and Ref.~\cite{kolodrubetz_16} for details):
\begin{equation}
\left[i\hbar \partial_\lambda H_0-\left[\mathcal A_\lambda,H_0 \right],H_0 \right]=0.
\label{eq:def_CDgauge}
\end{equation}
This equation immediately extends to classical systems by replacing the commutator with the Poisson brackets: $[\dots]\to i\hbar \{\dots\}$. While in this work we focus on quantum systems all general results and methodology equally apply to classical systems.

One can immediately see that Eq.~\eqref{eq:def_CDgauge} reduces to a familiar Galilean transformations in the case of translations. Indeed, let us assume that
\be
H_0={p^2\over 2m}+V(q-\lambda(t)),
\ee
i.e. $\lambda(t)$ is the position of the center of the potential. In this case the gauge potential is just the momentum operator $\mathcal A_\lambda=p$. Indeed we have $i\hbar \partial_\lambda H_0\equiv -i\hbar \partial_q V=-i\hbar [\partial_q, H_0]=[p,H_0]$ such that Eq.~\eqref{eq:def_CDgauge} is automatically satisfied. Thus the exact CD Hamiltonian is
\be
H_{\rm CD}=H_0+\dot\lambda p
\label{h_cd_tr}
\ee

If the waiter implements the protocol~\eqref{h_cd_tr} then he would be able to move a tray with glasses fast without exciting it. Notice that this is not what the waiter actually does. The reason is that it is very hard to realize the term linear in momentum. Moreover from Eq.~\eqref{eq:def_CDgauge} we see that in systems satisfying time-reversal symmetry, where the Hamiltonian $H_0$ is real, the gauge potential and hence the corresponding CD terms are always strictly imaginary. So the CD driving always breaks the time reversal symmetry. The situation is actually much better than it seems. The CD term in this setup plays the role of the vector potential in electro-magnetism:
\be
H_{\rm CD}={p^2\over 2m}+V(q-\lambda)+\dot\lambda p={(p+m\dot\lambda)^2\over 2m} +V(q-\lambda)-{m\dot\lambda^2\over 2}.
\ee
It is very well known that if the vector potential is curl-free, which is always the case in one dimension, it can be removed (gauged away) at the expense of introducing a scalar potential via $p\to p+\partial_q f$ and $H\to H-\partial_t f$ for an arbitrary function $f(q,t)$. Choosing $f(q,t)=-m\dot \lambda q$ we see that 
\be
H_{\rm CD}\sim {p^2\over 2m}+V(q-\lambda)+m\ddot\lambda q.
\label{h_cd_tr1}
\ee
Thus as expected from the Galilean invariance, the CD driving amounts to adding an extra gravitational field proportional to the acceleration. Here we used the tilde-sign instead of equal sign to highlight that the r.h.s. is gauge equivalent rather than equal to the Hamiltonian~\eqref{h_cd_tr}.  There is an important physical difference between the two CD protocols. While following the imaginary CD protocol~\eqref{h_cd_tr} amounts to instantaneously following eigenstates of $H_0$, following the real CD protocol~\eqref{h_cd_tr1} amounts to instantaneously following eigenstates of a gauge equivalent Hamiltonian. Only when the velocity $\dot\lambda$ is zero the two Hamiltonians coincide such that CD driving leads to identical results. This subtlety is precisely the reason why the waiter, following real CD protocol, does not keep the glass in the instantaneous ground state except when the velocity and the acceleration are zero.

Although Eq.~\eqref{eq:def_CDgauge} is linear and looks very simple, this simplicity is actually deceptive. In fact, one can show that in generic chaotic systems it has no solution. In quantum chaotic systems the exact analytic expression for $\mathcal A_\lambda$ suffers from the problem of small denominators~\cite{kolodrubetz_16} and in classical chaotic systems it can be expressed through a formally divergent integral~\cite{jarzynski_95}. The physical reason behind is very simple. By trying to find the exact gauge potential we are requiring too much. We essentially want to find a transformation which keeps the system in exact many-body eigenstates without any excitations. But this is clearly impossible or at least exponentially hard. E.g. in chaotic many-particle systems eigenstates are essentially random vectors sensitive to exponentially small perturbations of the Hamiltonian~\cite{dalessio_15} so the exact gauge potential should have the same exponential sensitivity to the details of the many-body spectra and have access to all microscopic degrees of freedom. But finding such gauge potential is hardly our goal either. We are generally either interested in suppressing dissipation, i.e. suppressing transitions between levels resulting in substantial energy changes, or in following very special states like the ground states, which are also robust to small perturbations. Thus our goal should be finding approximate gauge potentials, which satisfy requirements of robustness and locality and which strongly suppress physical diabatic effects rather than completely eliminate them. This is precisely what we are going to discuss next.

\subsection{Variational principle and local CD protocols}

%Let us introduce $\mathcal A_\lambda^\ast$ as an approximate adiabatic gauge potential, which satisfies the conditions like locality, robustness or experimental accessibility.

 Our goal is to set up a variational procedure allowing one to determine the best possible $\mathcal A_\lambda$ under some constraints like locality, robustness or just experimental accessibility. Unconstraint minimization should of course result in the exact Gauge potential \eqref{eq:def_CDgauge}.
It is easy to see that solving Eq.~\eqref{eq:def_CDgauge} is equivalent to minimizing the Hilbert-Schmidt norm of the operator
\[
G_\lambda(\mathcal A_\lambda) \equiv \partial_\lambda H_0+{i\over \hbar}[\mathcal A_\lambda,H_0]
\]
with respect to $\mathcal A_\lambda$. Ideed, finding the minimum of this norm is equivalent to the Euler-Lagrange equation
\begin{equation}
{\delta \mathcal S(\mathcal A_\lambda)\over \delta \mathcal A_\lambda}=0,
\label{eq:main}
\end{equation}   
of the action (see methods):
\begin{equation}
\mathcal{S}\left(\mathcal A_\lambda \right)=\tr \left[ G_\lambda^2(\mathcal A_\lambda) \right].
\label{eq:def_CDaction}
\end{equation}
For classical systems trace should be replaced by an integral over the phase space. Instead of minimizing the action over the whole Hilbert space of operators, one can now restrict to a subspace of physical operators. Let us introduce $\mathcal A_\lambda^\ast$ as an approximate adiabatic gauge potential, then one can simply calculate the best constrained approximation by minimizing over all allowed $\mathcal A_\lambda^\ast$. To do so one simply has to be able to evaluate the action \eqref{eq:def_CDaction} for the allowed operators $\mathcal A_\lambda^\ast$. Evaluating the trace of local operators or their products is very straightforward and can be usually done with minimal efforts. Physically, the action~\eqref{eq:def_CDaction} defines the average transition rate (over all possible states) when the classical parameter $\lambda(t)$ is a weak random white-noise process. So minimizing the action is equivalent to suppressing  processes such as heating and energy diffusion under a white noise drive.

Quite often one is interested in suppressing transitions from a low-temperature manifold of states, in particular, from the ground state. Then targeting the gauge potential, which suppresses transitions everywhere in the spectrum is over-demanding. Instead one can define the action through a finite temperature norm:
\begin{equation}
\mathcal{S}\left(\mathcal A^\ast_\lambda,\beta \right)=\left< G_\lambda^2(\mathcal A_\lambda^\ast)\right>-\left< G_\lambda(\mathcal A_\lambda^\ast)\right>^2,
\label{eq:def_CDaction1}
\end{equation}
where the angular brackets denote usual average with respect to the thermal density matrix $\rho(\beta)={1\over Z}\exp[-\beta H_0]$. The Hilbert-Schmidt norm~\eqref{eq:def_CDaction} is recovered as the infinite temperature limit $(\beta\to 0)$ of this norm (note that subtracting the second term in Eq.~\eqref{eq:def_CDaction1} is not affecting the result as $\langle G_\lambda(\mathcal A_\lambda^\ast)\rangle$ is independent of $\mathcal A_\lambda^\ast$). In the zero temperature limit $\beta\to\infty$ the action~\eqref{eq:def_CDaction1} reduces to the variance of $G_\lambda$ in the ground state. The exact gauge potential minimizes the action~\eqref{eq:def_CDaction} for any temperature. However, details of the variational solution can depend on $\beta$. In this work we are focusing on finding CD protocols minimizing the infinite-temperature action~\eqref{eq:def_CDaction} leaving the analysis of the finite/zero temperature action for future work.

\section{Applications}
From this point on we will only be concerned with quantum systems and will set $\hbar=1$.
\subsection{One-dimensional lattice fermions}

Let's investigate the performance of variational CD protocols using an example of non-interacting lattice spinless fermions in a time-dependent potential. Despite the simplicity of the setup, the problems we will be addressing are highly nontrivial as we will consider inserting and moving obstacles breaking translational symmetry. In turn, this generally leads to strong mixing of single-particle orbitals and strong non-adiabatic effects originating from scattering of fermions from the obstacle. Consider a single-band tight binding model in an external potential
\be
H_0=-J\sum_{j=1}^{L-1} (c_j^\dagger c_{j+1}+c_{j+1}^\dagger c_j)+\sum_{j=1}^L V_{j}(\lambda) c_j^\dagger c_j,
\label{eq:varA}
\ee
where $c_j^\dagger$ and $c_j$ are fermionic creation and annihilation operators and $V_j(\lambda)$ is the external potential. In principle the dependence of $V$ on $\lambda$ can be arbitrary. Here we will focus on a particular example of inserting the potential: $V_j(\lambda)=\lambda v_j$. In supplementary information we additionally analyze a moving potential: $V_j(\lambda)=V(j-\lambda)$. Because the Hamiltonian is real, as we discussed earlier, the adiabatic gauge potential should be purely imaginary (this is also clear from the form of the action~\eqref{eq:def_CDaction}). The most local Hermitian imaginary operator one can find for a free system is the current, therefore we will look for solutions of the form: 
\[
\mathcal A_\lambda^\ast=i \sum_{j=1}^{L-1} \alpha_j (c_{j+1}^\dagger c_j-c_j^\dagger c_{j+1}).
\]
Substituting this potential into the action~\eqref{eq:def_CDaction} and extremizing with respect to the coefficients $\alpha_j$ we find the following equation (see supplementary information)
\be
-3 J^2 \partial^2_j \alpha_j+(\partial_j V_j(\lambda))^2 \alpha_j=-J\partial_j \partial_\lambda V_j(\lambda),
\label{eq:varA}
\ee
where $\partial_j \alpha_j\equiv \alpha_{j+1}-\alpha_j$, $\partial_j V_j\equiv V_{j+1}-V_j$ and $\partial^2_j \alpha_j\equiv \alpha_{j+1}-2\alpha_j+\alpha_{j-1}$ are discrete lattice derivatives. If the potential is smooth on the scale of the lattice spacing then the discrete derivatives can be substituted by the continuous derivatives. In general, this is a set of linear equations, which can always be solved numerically. However, there are several cases where one can find approximate or exact analytic solutions. 

The imaginary hopping terms in $\mathcal A_\lambda^\ast$ explicitly break time reversal symmetry, which might create some difficulties in implementing them in practice. However, restricting the variational ansatz only to the nearest neighbors allows one once again to perform a simple gauge transformation similar to the vector potential shift also known as the Peierls: substitution. Namely, $c_j\to c_j \mathrm e^{-i f_j}$ such that the new Hamiltonian becomes real (see supplementary information for more details):
\be
H_{CD}\sim -\sum_j J_j \left( c^\dagger_{j+1} c_j +h.c. \right)+ \sum_j U_j  c^\dagger_{j} c_j,
\label{eq:H_CD_1Dfermions}
\ee
with 
\begin{eqnarray}
J_j&=&J\sqrt{1+(\dot{\lambda} \alpha_j/J )^2}, \\
U_j&=&V_{j}-\sum_{i=1}^j \frac{J}{J^2+(\dot{\lambda}\alpha_j)^2} \left(\ddot{\lambda} \alpha_j+(\dot{\lambda})^2\partial_\lambda \alpha_j \right).
\end{eqnarray}
As earlier the tilde sign indicates that the r.h.s. is gauge equivalent to the CD Hamiltonian and the two coincide only at $\dot\lambda=0$. This is again very similar to the tilting the tray in the waiter example. It is easy to see that the CD Hamiltonian has two distinct limits: At small velocities $|dV_j/dt|=|\dot\lambda v_j|\ll J^2$ the renormalization of hopping in negligible and the CD term is a correction to the potential proportional to the acceleration $\ddot\lambda$ exactly like in the waiter example. Conversely in the high velocity limit renormalization of the potential is negligible and the CD Hamiltonian contains the renormalized hopping, which scales linearly with the velocity. This local hopping renormalization plays a role similar to the refractive index by locally changing the group velocity of electrons in a way, which essentially traps scattered electrons.

As the CD Hamiltonian depends only on the velocity and acceleration $\dot\lambda$ and $\ddot\lambda$ it is convenient (though not necessary) to deal with $\lambda(t)$ which have vanishing first and second derivatives in the beginning and the end of the protocol such that $H_{CD}=H_0$ at these points. An example of such a protocol, we will use in this work is
\be
\lambda(t)=\lambda_0+(\lambda_f-\lambda_0)\, \sin^2\left({\pi\over 2} \sin^2\left( {\pi t\over 2\tau}\right)\right),\quad t\in (0,\tau).
\label{eq:lambda_t}
\ee
This protocol ramps $\lambda(t)$ from the initial value $\lambda_0$ to the final value $\lambda_f$ during the time $\tau$.

{\em Uniform linear potential.} In the case of a general time-dependent force $V_j(\lambda)=\lambda j$, with $\lambda$ playing the role of an effective electric field, the solution of Eq.~\eqref{eq:varA} is very simple: $\alpha_j=-J/\lambda^2$. It is easy to check that this solution is in fact exact up to the boundary terms (see supplementary material for details), i.e. $\mathcal A_\lambda^\ast=\mathcal A_\lambda$. Remarkably, because $\alpha$ is constant, the effective potential $U_j$ remains linear. Additionally, the effective hopping is constant across the lattice, which allows us to absorb the hopping renormalization into the timescale, as one can always rescale the Hamiltonian by an arbitrary factor without exciting the system. 

\begin{figure}[t]
  \centering
  \includegraphics[width=\linewidth]{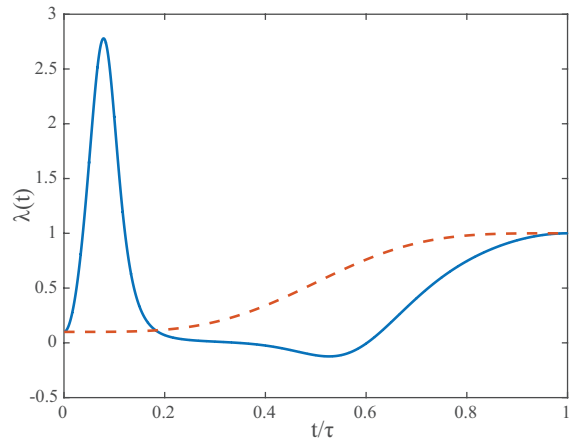}
  \caption{{\bf CD Protocol.} Example of a naive adiabatic protocol \eqref{eq:lambda_t} to switch on a linear potential from $\lambda_0=0.1J$  to a final value of $\lambda_f=J$ without exciting the system (dahsed red) and a counter diabatic protcol that does exactly that in a time $\tau=5/J$ (full blue). }
  \label{fig:electric field}
\end{figure}

As a result of these transformations the real CD Hamiltonian is structurally the same as the naive Hamiltonian. One simply has to switch on the electric field in a different way to avoid ending in an excited state, i.e. for each protocol $\lambda$ the CD protcol is
\begin{equation}
\lambda_{CD}=\frac{\lambda}{\sqrt{1+\dot{\mu}^2}}\left(1-\frac{\mu \ddot{\mu}}{1+\dot{\mu}^2} \right) \quad{\rm where}\; \mu=1/\lambda.
\end{equation}
This protocol is illustrated in Fig.~\ref{fig:electric field} for $\lambda_0=0.1J$ and $\lambda_f= J$ (full blue line) for a particular choice $\tau=5/J$. 
In order to allow for enough transport of particles there is an initial pulse in the field with an amplitude that is significantly larger than the final one. After this pulse the field is much flatter and first goes opposite to the target field before reversing to the right direction to reach the final desired field $\lambda_f$. The naive and the CD protocols come closer to each other as one increases the duration $\tau$, though they always significantly differ at small fields.

Let us also comment on the opposite limit of instantaneous protocol, $\tau\to 0$, where the bare coupling turns on as a step-like function: $\lambda(t)\to\lambda_0+(\lambda_f-\lambda_0)\theta(t)$. In this case the Galilean term dominates the CD Hamiltonian. It is then convenient to formally use the chain rule and parametrize time in terms of the coupling: $i\partial_t \psi=i\dot\lambda \partial_\lambda \psi$ such that the Schr\"odinger equation reads:
\be
i\partial_\lambda |\psi\rangle=\mathcal A_\lambda^\ast |\psi\rangle
\label{eq:CD_fast}
\ee
If the gauge potential is exact then this equation describes the fastest route to perform adiabatic evolution. In particular, a slow turning on of the uniform field is equivalent to dynamics governed by the Hamiltonian
\be
\mathcal A_\lambda^\ast=-{i J\over \lambda^2}\sum_j (c_{j+1}^\dagger c_j-c_j^\dagger c_{j+1}).
\ee
Since the coupling $\lambda$ effectively plays the role of time, we see that the total time required to load the system into the ground state according to  Eq.~\eqref{eq:CD_fast} diverges as the initial or final electric field approaches zero. As we will discuss elsewhere this divergence is fundamental related to the divergence of the quantum speed limit~\cite{sels_in_prep}. Let us point that at $\tau\to 0$ there is no smooth real CD protocol as it contains the acceleration terms which become singular.

{\em Fighting Anderson orthogonality: Inserting a potential.} A second, somewhat more involved problem is the adiabatic insertion of a scattering potential into the Fermi-gas. This problem is 
harder than it might seem. The difficulty can be understood from the the perspective of Anderson's orthogonality catastrophe~\cite{anderson}, which states that the ground state of the homogeneous Fermi-gas and the gas with a single impurity are orthogonal in the thermodynamic limit. In addition the system is gapless, as a consequence standard arguments exploited in the adiabatic quantum computing literature~\cite{farhi_00, altshuler_10} suggest that, in order to load the potential adiabatically, one has to scale the ramp velocity to zero with the inverse system size. We checked that this is indeed the case for the naive loading protocol. The situation changes dramatically with the CD driving.  

\begin{figure}[b]
  \centering
  \includegraphics[width=\linewidth]{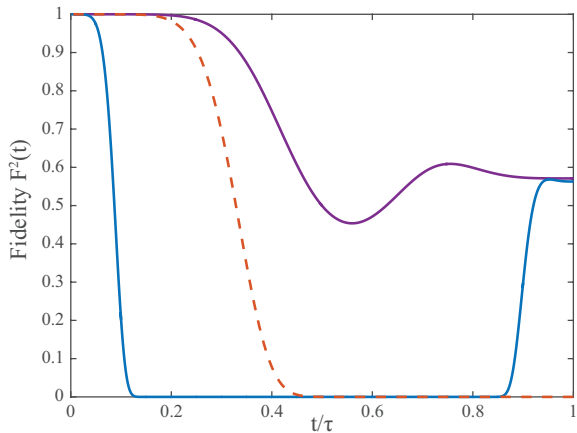}
  \caption{{\bf Inserting local potential.} The probability to be in the adiabatic ground state when inserting a scattering potential quickly decays to a small value for the naive protocol (dashed red line). By CD driving with a local complex gauge, the state stays much closer to the ground state and a final fidelity of about $1/2$ is reached (full purple line). A gauge equivalent real Hamiltonian, with renormalized hopping and potential, results in the same final fidelity but is almost orthogonal to the ground state at intermediate times (full blue line).   }
  \label{fig:insert}
\end{figure}

To obtain the gauge potential we numerically solve Eq.~\eqref{eq:varA} in a box of size $L$ with vanishing boundary conditions $\alpha_{j=-L/2}=\alpha_{j=L/2}=0$. 
In Fig.(\ref{fig:insert}) we show time dependence of the squared fidelity of the wave function and the instantaneous ground state: $F^2(t)=\left| \langle \psi(t)|\psi_{GS}(t)\rangle\right|^2$ for different protocols. The total system size is $L=512$ at half filling (256 particles). The system is initially prepared in the ground state at zero potential and then we are turning on a repulsive Eckart potential of the form:
\[
V_j(\lambda)={\lambda\over \cosh^2 j/\xi},\; j\in[-L/2,L/2]
\]
We choose $\xi=8$ and turn on $\lambda(t)$ according to the protocol~\eqref{eq:lambda_t} with $\lambda_0=0$ and $\lambda_f=2J$. The total duration of the protocol is $\tau=10/J$. The naive protocol indeed fails completely, giving the final fidelity $F^2(\tau)\approx 2\cdot 10^{-19}$ as expected. This is only marginally better than the fidelity of the initial state and final state which is $5\cdot 10^{-20}$. The CD protocol on the other hand gives fidelity of the order of $1/2$ gaining more than 18 orders of magnitude. This value implies a $50\%$ chance of preparing the system in the exact many-body ground state. Notice that while the imaginary CD protocol (solid purple line) keeps instantaneous fidelity high at all times, for the real protocol, exactly like in the waiter case, the instantaneous fidelity at intermediate times drops to a very small value. High fidelity is only recovered at the end of the protocol, where the velocity $\dot\lambda$ becomes close to zero.

\subsection{One-dimensional spin chain}

In the previous examples we focused on free-particle one-dimensional systems, where an exact solution for $\mathcal A_\lambda$ always exists and the variational approach merely helps one to find a simpler and easier to implement local approximate $\mathcal A_\lambda^\ast$. Many-particle systems are intrinsically chaotic and as we already discussed, the exact gauge potential simply does not exist in the form of a local operator. In such cases approximate methods for finding $\mathcal A_\lambda^\ast$ are simply required to find CD protocols. To illustrate the power of the variational approach let us consider an Ising spin chain in the presence of a transverse and longitudinal field. This is one of the simplest non-integrable models with very rich phase diagram (see e.g. Refs.~\cite{simon_11, kim_13}). The Hamiltonian of this system reads
\be
H_0=\sum_j \left(J \sigma_j^z\sigma_{j+1}^z +Z_j \sigma_j^z + X_j \sigma_j^x \right),
\label{eq:spinH}
\ee
where $\sigma_j^z$ and $\sigma_j^x$ are the Pauli matrices. We allow all couplings to depend on some tuning parameter $\lambda$, which in turn depends on time. The simplest gauge potential, which is purely imaginary, is the magnetic field along the y-direction:
\be
\mathcal A_\lambda^\ast=\sum_j \alpha_j \sigma_j^y,
\label{eq:CDspinint0}
\ee
Recall that $G_\lambda \equiv \partial_{\lambda} H+i [\mathcal A_\lambda^\ast, H]$, hence
\begin{eqnarray}
 G_\lambda&=&\sum_j \left( (X_j'-2 Z_j \alpha_j) \sigma^x_j+ 2\alpha J (\sigma^x_j \sigma^z_{j+1}+\sigma^z_j \sigma^x_{j+1}) \right) \nonumber \\
 &&+\sum_j \left( (Z_j'+2X_j \alpha_j) \sigma^z_j+J'  \sigma_j^z\sigma_{j+1}^z \right),
\end{eqnarray}
where `prime' stands for the derivative with respect to $\lambda$. Since Pauli matrices are traceless, computing the Hilbert-Schmidt norm of this operator is trivial and amounts to adding up squares of coefficients in front of independent spin terms:
\begin{eqnarray}
2^{-L}\tr(G_\lambda^2)&=&\sum_j \left((X_j'-2 Z_j \alpha_j)^2+8\alpha_j^2 J^2 \right) \nonumber \\
&&+\sum_j \left((Z_j'+2X_j \alpha_j)^2+(J')^2 \right).
\end{eqnarray}
Minimizing the quadratic form with respect to $\alpha_j$ immediately yields the optimal variational solution:
\be
\alpha_j= {1\over 2}{Z_j  X'_j-X_j Z'_j \over Z_j^2+X_j^2+2J^2},
\label{eq:alpha_q}
\ee
Note that the y-magnetic field is strictly local as it only depends on the local values of the x and z magnetic field. For $J=0$ this gauge potential is exact, as it is simply a generator of local spin-rotations in $x-z$ plane. Note that $\mathcal A_\lambda^\ast$ vanishes if either $h_x=0$ or $h_z=0$ implying that the leading contribution to $\mathcal A_\lambda$ in this case actually comes from two-spin terms. To second order we can include two-spin terms into the variational ansatz:

\beq
\mathcal A_\lambda^\ast=\sum_j \left[ \alpha_j \sigma_j^y+ \beta_j \left( \sigma_j^y\sigma_{j+1}^z+ \sigma_j^z\sigma_{j+1}^y \right) \right] \nonumber \\
+\sum_j\gamma_j \left( \sigma_j^y\sigma_{j+1}^x+ \sigma_j^x\sigma_{j+1}^y \right).
\label{eq:CDspinint}
\eeq
The coefficients $\alpha_j,\beta_j,\gamma_j$ can directly be found by minimizing the quadratic action \eqref{eq:def_CDaction} resulting in a linear set of coupled equations, which can be easily solved numerically. This variational solution dramatically enhances performance of the annealing protocol loading spins from an initial product state to the ground state of the Hamiltonian \eqref{eq:spinH} (see Sup. Info).

\begin{figure}
  \centering
  \includegraphics[width=\linewidth]{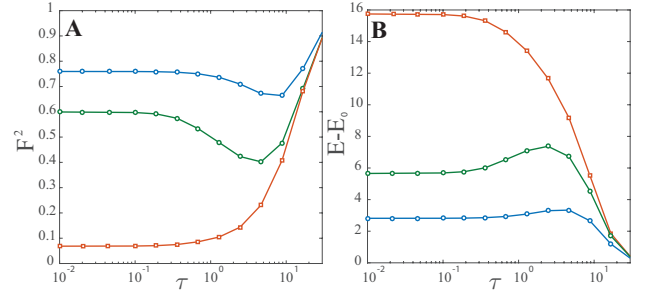}
  \caption{{\bf Local spin flip.} A single spin in a chain of 15 spins is aligned with the x-direction by switching a strong local magnetic field in $x$-direction (see text for details) . Panel {\bf A} shows the fidelity to recover the system in the ground state ofter a protocol of duration $\tau$. Panel {\bf B} shows the final energy above the ground state. The red squares are associated with the naive protocol, the green circles with strictly local CD driving and the blue with CD driving \eqref{eq:CDspinint}. }
  \label{fig:spinchainGS}
\end{figure}

Following the orthogonality catastrophe example, let us consider a CD protocol for turning on a local magnetic field. Specifically we consider turning on an additional $x$-magnetic field $\lambda$ from zero to the final value $\lambda_f=-10J$ in a periodic chain described by the Hamiltonian $H_0+\lambda \sigma_0^x$, where $H_0$ is given by Eq.~\eqref{eq:spinH} with $J=1$, $Z_j=2$, $X_j=0.8$. We compare three protocols: bare protocol with no CD driving, CD protocol with only local CD driving (Eq.~\eqref{eq:CDspinint0}) and the optimal two-spin CD protocol (Eq.~\eqref{eq:CDspinint}). We verified that the variational coefficients $\alpha_j,\beta_j,\gamma_j$ rapidly (exponentially) decay with $j$ away from the site $j=0$ so the CD protocol effectively remains local. Note that we can find the CD protocol in the thermodynamic limit, however, we have to apply it to a finite chain to verify its performance. The probability of recovering the system in the ground state for a chain of $15$ spins is shown in Fig.~\ref{fig:spinchainGS}. For fast protocols there is a significant reduction in excess energy and a corresponding increase in fidelity for the counter-diabatic protocols. In particular, in the instantaneous quench limit $\tau\to 0$ the CD driving gives almost a factor of ten gain in fidelity and a similar reduction in the heating. Interestingly in this limit dynamics is entirely governed by $\mathcal A_\lambda^\ast$, which is exponentially localized near $j=0$. Loading into a strong magnetic field can be also viewed as changing the boundary conditions in the system, effectively it cuts the chain in two pieces. Moreover, due to time reversal symmetry the fidelity of joining the chain back together is identical. So in this case $\mathcal A_\lambda^\ast$ can be interpreted as an effective boundary Hamiltonian generating adiabatic boundary transformations on the ground state wave function.

\begin{figure}
  \centering
  \includegraphics[width=\linewidth]{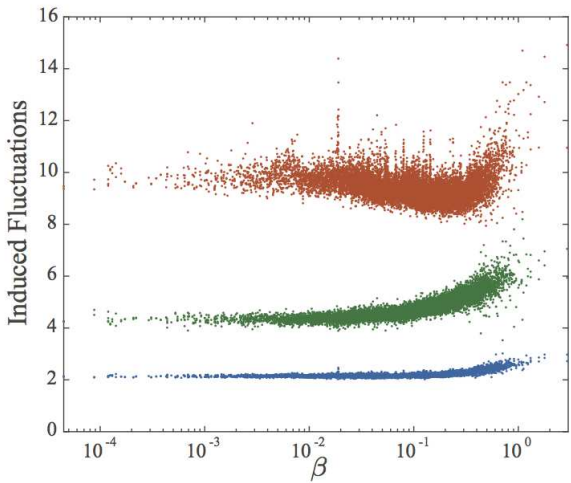}
  \caption{{\bf Spin chain fluctuations.} Normalized energy fluctuations production rate in a chain of $L=15$ spins with $X=0.9$, $Z=0.8$ and $J=1$, when a single spin is subject to a random weak magnetic field in the $x$-direction. The effective inverse temperature of each eigenstate $\beta$ is defined to match its energy in an equivalent thermal ensemble. The red, green and blue dots represent results for no CD term, best single site CD term and the best two-site CD term respectively.}
  \label{fig:spinchainFluct}
\end{figure}

Next let us use the same example to analyze another application of the CD driving, namely suppression of dissipation in a noisy system. Let us now assume that the site $j=0$ is subject to a small white noise in the magnetic field in the $x$-direction. Physically this noise can stem from coupling of this site to a nearby impurity or a quantum dot. A convenient measure of dissipation in the system is the rate of spread of energy fluctuations $d_t \delta_E^2$. If we implement exact CD protocol then the system follows instantaneous eigenstates and the energy fluctuations remain constant in time so the rate is zero. At finite temperatures $d_t \delta_E^2$ is related to the usual heating rate $dE/dt$ by the fluctuation dissipation relation (see e.g. Ref.~\cite{dalessio_15}). However, the spread of energy fluctuations has a well defined limit even for the infinite temperature states, where the heating rate vanishes. Within Fermi's golden rule, and under the assumption that the spectral density of the fluctuating field is white, the rate at which the fluctuations will increase if we initialize the system in a pure state $\left|n\right\rangle$, takes the following form:
\begin{equation}
\partial_t \delta_E^2 = S_{\lambda \lambda} \left\langle n \right| \left( \left[H_0,G_\lambda\right] \right)^2 \left| n \right\rangle,
\end{equation}
where $S_{\lambda\lambda}$ is the power spectral density of the noise, i.e. the inverse bandwidth of the noise. 
Figure \ref{fig:spinchainFluct} shows the normalized production rate $\partial_t \delta_E^2/ S_{\lambda \lambda}$ for every eigenstate of an ergodic chain of 15 spins. We clearly see a reduction in fluctuations across the entire spectrum, comparable to the reduction in excess energy for the ground state protocol discussed above. Interestingly 
CD driving not only reduces dissipation but also reduces its fluctuations between different eigenstates.

\begin{figure}
  \centering
  \includegraphics[width=\linewidth]{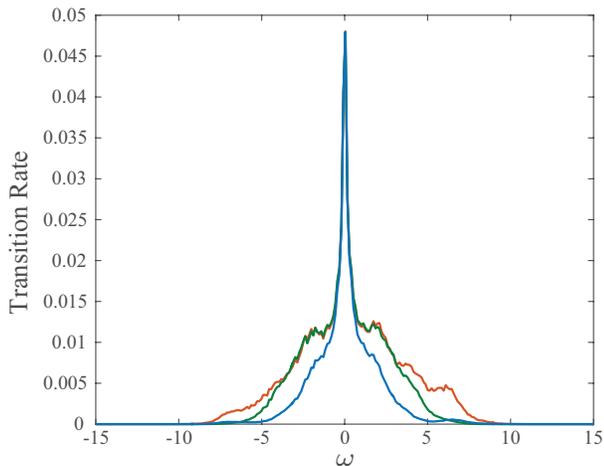}
  \caption{{\bf Spin chain transitions.} Average normalized transition rate over states with effective temperature $\beta=0.1$. The parameters and the colors are the same as in Fig.~\ref{fig:spinchainFluct}.}
  \label{fig:spinchainTrans}
\end{figure}

To understand better the performance of the CD protocol we can analyze the spectral decomposition of the dissipation as a function of the absorption frequency. Specifically we look at the lifetime of a state $\Gamma_n$, within Fermi's golden rule, when we subject the system to a periodic drive with frequency $\omega$
\begin{equation}
\Gamma_n(\omega)= S_{\lambda \lambda} (\omega)  \sum_m \left|\left\langle m \right| G_\lambda  \left| n\right\rangle\right|^2 \delta(E_m-E_n-\omega),
\end{equation} 
where the $\delta$-function is broadened such that it contains several eigenstates leading to a smooth dependence of $\Gamma_n(\omega)$. The result is shown in Fig.~\ref{fig:spinchainTrans}. All three protocols show a very narrow peak in the transition rate around $\omega=0$, followed by a much broader distribution at large frequencies. The small frequency transitions represent mixing between nearby eigenstates. Those lead to very small dissipation and are not affected by the CD driving terms. Conversely the high frequency transitions leading to large energy transfer from the noise to the system and hence to dissipation are strongly suppressed by the CD driving. 

\subsection{Classical Spin Chain}
Let us briefly show how the developed ideas can be applied to classical systems. Specifically we will consider a classical rotor model with the Hamiltonian 
\be
H_0=J\sum_j \vec S_j \vec S_{j+1}+X_0(\lambda S_0^x+ Z_0(\lambda) S_0^z,
\ee
where $\vec S_j$ are three-dimensional classical angular momenta. Because $\vec S_j^2=S^2$ is conserved under the Hamiltonian dynamics we will set $S^2=1$. This Hamiltonian, with appropriate rescaling of the coupling constants, can be obtained as a large spin limit of a quantum Heisenberg spin chain with an additional local magnetic field. This Hamiltonian is a direct analogue of a quantum spin model analyzed earlier~\eqref{eq:spinH} except that we use isotropic Heisenberg coupling. Because the Heisenberg model is not integrable there is no need to introduce extra static magnetic fields. As in the quantum case in the leading order we will seek the gauge potential in the form
\be
\mathcal A_\lambda^\ast=\alpha_0(\lambda) S_0^y.
\ee
Recall that the Poisson bracket between any two functions can be expressed as
\[
\left\{A(\vec S_j),B(\vec S_j)\right\}=\sum_j \epsilon_{abc} S_j^a {\partial A\over \partial S_j^b}{\partial B\over \partial S_j^c},
\]
where $a,b,c=\{x,y,z\}$ and $\epsilon_{abc}$ is a fully antisymmetric tensor. Using this we find that
\begin{multline}
G_\lambda=\partial_\lambda H_0+\left\{H_0,\mathcal A_\lambda^\ast\right\}
=(X_0'-\alpha Z_0) S_0^x+(Z_0'+\alpha X_0) S_0^z\\
+J\alpha \left(S_0^z(S_1^x+S_{-1}^x)-S_0^x(S_1^z+S_{-1}^z)\right)
\end{multline}
From this we easily obtain $||G_\lambda||^2$ by integrating the square of $G_\lambda$ over spin directions:
\be
||G_\lambda||^2={1\over 3}(X_0'-\alpha Z_0)^2+{1\over 3} (Z_0'+\alpha X_0)^2+{4\over 9} J^2\alpha^2.
\ee
Minimizing this with respect to $\alpha$ we find the optimum solution:
\be
\alpha={Z_0 X_0'-X_0 Z_0'\over X_0^2+Z_0^2+{4\over 3} J^2}.
\ee
This expression is very similar to the quantum result~\eqref{eq:alpha_q} with a slightly different prefactor in the $J^2$ (The overal factor 1/2 in front is just due to the difference between spins and Pauli matrices). For the classical model with only $z-z$ interactions the only difference in $\alpha$ would be in a prefactor in front of $J^2$: $2/3$ instead of $4/3$. In a similar fashion one can extend the variational ansatz to higher orders including various terms odd in powers of $S^y$ like $S_j^y S_{j+1}^{(z,x)}$.

\section{Methods}

Consider a state $\left|\psi\right\rangle$, evolving under a time dependent Hamiltonian $H_0(\lambda(t))$
\begin{equation}
i \hbar\partial_t \left|\psi\right\rangle= H_0(\lambda(t)) \left|\psi\right\rangle,
\end{equation}
where the full time dependence of the Hamiltonian is caused by an external drive $\lambda(t)$. Let us go to the rotating frame where the Hamiltonian remains stationary (diagonal at all times). This can be always achieved by a $\lambda$-dependent unitary transformation $U(\lambda(t))$ which also expresses the state in the instantaneous eigenbasis of the Hamiltonian. It is easy to see that the wave function in the moving frame is $|\tilde \psi \rangle=U(\lambda)|\psi \rangle$ satisfies the effective Schr\"odinger equation~\cite{kolodrubetz_16}
 \begin{equation}
i \hbar\partial_t |\tilde\psi\rangle=\left( \tilde H_0(\lambda(t))-\dot{\lambda}\tilde{\mathcal A}_\lambda\right)|\tilde \psi\rangle
\end{equation}
where $\tilde H_0$ is the Hamiltonian in the instantaneous basis and $\tilde {\mathcal A}_\lambda$ is the adiabatic gauge potential in the moving frame: 
\begin{eqnarray}
&&\tilde H_0(\lambda(t))=U^\dagger H_0(\lambda(t)) U= \sum_n \epsilon_n(\lambda) | n\rangle \langle  n|,\nonumber\\ 
&&\tilde {\mathcal A}_\lambda=i U^\dagger \partial_\lambda U
\end{eqnarray}
Differentiating the relation $\tilde H_0=U^\dagger H_0 U$ with respect to $\lambda$ and using the fact that $\partial_\lambda \tilde H_0$ commutes with $\tilde H_0$ one can check that the adiabatic gauge potential satisfies the following equation~\cite{kolodrubetz_16}:
\begin{equation}
i\hbar \left( \partial_\lambda H_0+F_{\rm ad} \right)=\left[\mathcal A_\lambda,H_0 \right],
\label{eq:AG_eqmot}
\end{equation}
where $F_{\rm ad}$ is the adiabatic or generalized force operator:
\begin{equation}
F_{\rm ad}=-\sum_n \partial_\lambda\epsilon_n(\lambda) \left|n(\lambda)\right\rangle \left\langle n(\lambda) \right|.
\end{equation} 
and $\mathcal A_\lambda=U\tilde {\mathcal A}_\lambda U^\dagger=i (\partial_\lambda U)U^\dagger$ is the adiabatic gauge potential in the lab frame. While we used the moving frame to derive Eq.~\eqref{eq:AG_eqmot}, it is an operator equation, which is valid in any frame including the lab frame. Note that Eq.~\eqref{eq:def_CDgauge} trivially follows from Eq.~\eqref{eq:AG_eqmot} because $F_{\rm ad}$ by construction commutes with the Hamiltonian.

Now let us discuss in more detail how Eq.~\eqref{eq:AG_eqmot} can be reformulated as the minimization problem leading to the variational ansatz. For this purpose let us choose some trial gauge potential $\mathcal A_\lambda^\ast$ and define an operator $G_\lambda$:
\[
G_\lambda=\partial_\lambda H_0+{i\over \hbar}[\mathcal A_\lambda^\ast, H_0].
\]
This operator also has a well-defined classical limit. It is clear from Eq.~\eqref{eq:AG_eqmot} that for $\mathcal A_\lambda^\ast=\mathcal A_\lambda$ we have $G_\lambda=-F_{\rm ad}$. The diagonal elements of $G_\lambda$ in the basis of the Hamiltonian do not depend on $\mathcal A_\lambda^\ast$: $\langle n|G_\lambda |n\rangle=\partial_\lambda \epsilon_n(\lambda)$. Thus different choices of $\mathcal A_\lambda^\ast$ only affect the off-diagonal elements of $G_\lambda$. The true gauge potential has no off-diagonal elements and thus corresponds to the operator $G_\lambda$ with the minimum Hilbert-Schmidt norm. Formally this can be seen from the distance between $G_\lambda$ and $-F_{\rm ad}$: 
\begin{multline}
\mathcal{D}\left(\mathcal A^\ast_\lambda\right)=\tr \left[\left(G_\lambda+F_{\rm ad}\right)^2\right]\\
=\tr \left[ \left(\partial_\lambda H_0+F_{ad}+\frac{i}{\hbar}\left[\mathcal A^\ast_\lambda,H_0 \right] \right)^2 \right],
\end{multline}
Using cyclic properties of the trace, this distance becomes
\begin{equation}
\mathcal{D}\left(\mathcal A^\ast_\lambda\right)=-\tr \left[F_{\rm ad}^2 \right]+\mathcal{S}\left(\mathcal A^\ast_\lambda\right),
\end{equation}
with the action $\mathcal{S}(\mathcal A_\lambda^\ast)$ given by Eq.~\eqref{eq:def_CDaction}. Minimizing the distance with respect to $\mathcal A_\lambda^*$ results in Eq.~\eqref{eq:def_CDgauge}.  We'd like to stress that an enormous gain has been made by moving from the original equation \eqref{eq:AG_eqmot} to \eqref{eq:def_CDgauge} because the adiabatic force has been eliminated. In fact, neither the action \eqref{eq:def_CDaction} nor the Euler-Lagrange equation \eqref{eq:def_CDgauge} make any reference to the adiabatic force, which is generally hard to compute.

\section{Discussion and Conclusion}
Building on the concept of transitionless driving we have developed a variational principle that allows one to construct approximate variational CD protocols. Using this variational ansatz we obtained best local CD protocols and demonstrated that they can strongly decrease dissipation in highly excited states and increase fidelity of the ground state preparation by many orders of magnitude. Efficient CD protocols can find many different applications from constructing fast and efficient annealing protocols both for quantum computers and quantum simulators to engineering thermodynamic engines operating close to the maximum efficiency at fast speeds. A key advantage of the variational method is that it allows one to find such protocols without need of knowing any details about the spectrum of the Hamiltonian or the structure of its eigenstates, which are usually very hard to obtain.

We illustrated the ideas using various examples: (i) inserting (and moving) a local potential barrier into a Fermi sea effectively fighting the Anderson orthogonality catastrophe, (ii) locally flipping a spin an ergodic quantum spin chain. In all cases we showed that CD gives dramatic decrease of dissipation and improvements in the final fidelity even in the gapless regimes where standard arguments based on the gap indicate that the high fidelity state preparation is not possible at these fast rates. It remains to be seen how efficient and robust counter-diabatic protocol can be in more complicated situations such as crossing phase transitions or annealing systems with slow glassy dynamics. 

\acknowledgements
We acknowledge useful discussions with C. Ching and C. Jarzynski. D.S. was supported by the FWO as post-doctoral fellow of the Research Foundation - Flanders. A.P. was supported by AFOSR FA9550-16-1-0334, NSF DMR-1506340 and ARO W911NF1410540.

\pagebreak
\widetext
\begin{center}
\textbf{\large Supplemental Information: Minimizing irreversible losses in quantum systems by local counter-diabatic driving}
\end{center}
\begin{center}
Dries Sels, Anatoli Polkovnikov \\
\emph{Department of Physics, Boston University, MA 02215, USA}
\end{center}
\setcounter{equation}{0}
\setcounter{figure}{0}
\setcounter{table}{0}
\setcounter{page}{1}
\renewcommand{\theequation}{S\arabic{equation}}
\renewcommand{\thefigure}{S\arabic{figure}}
\renewcommand{\bibnumfmt}[1]{[S#1]}
\renewcommand{\citenumfont}[1]{S#1}

This note provides supporting calculations and additional results to the paper \emph{"Minimizing irreversible losses in  quantum systems by local counter-diabatic driving"}. The material is separated in two sections. The first section deals with free fermions and the second with Ising type spin chains. Throughout this note we set $\hbar=1$.
\section{Optimal local counter-diabatic gauge for free fermions}
Here we derive the optimal local counter-diabatic gauge for transitionless driving of a general free fermion problem. We will be focus on the Hamiltonians of the form 
\begin{equation}
H_0=-J\sum_j \left( c^\dagger_{j+1} c_j +h.c. \right)+ \sum_j V_j(\lambda)  c^\dagger_{j} c_j,
\end{equation}
where $c^\dagger_j$ creates a fermion on site $j$ and $c_j$ annihilates the fermion. Recall that the approximate adiabatic gauge potential defining CD driving should minimize the following action (Eq. (8) from the main text):
\begin{equation}
\mathcal{S}\left(\mathcal A^\ast_\lambda\right)={\rm Tr}\left[G_\lambda^2\right],
\label{eq:def_CDaction}
\end{equation}
where 
\begin{equation}
G_\lambda=\partial_\lambda H_0+i\left[\mathcal A^\ast_\lambda,H_0 \right] 
\end{equation}
For quadratic problems the adiabatic gauge potential is also quadratic. Because it is also imaginary it has to be expressed in the form:
\be
\mathcal A^\ast_\lambda=i \sum_{j,k} \alpha_{j,k}\left( c^\dagger_{k} c_j -h.c.\right),
\ee
where $\alpha_{j,k}=-\alpha_{k,j}$ and all elements are real. In this work we are not concerned with finding exact adiabatic gauge potentials but rather in their best local approximations. Thus, as in the main text, we are restricting $\mathcal A^\ast_\lambda$ to the following form
\begin{equation}
\mathcal A^\ast_\lambda=i \sum_j \alpha_j\left( c^\dagger_{j+1} c_j -h.c.\right),
\end{equation}
and treat coefficients $\alpha_j$ as variational parameters. It is straightforward to check that
\be
G_\lambda=\sum_j \left(\partial_\lambda V_j - 2J (\alpha_j-\alpha_{j-1})\right) c_j^\dagger c_j+J\sum_j (\alpha_{j}-\alpha_{j-1})(c_{j+1}^\dagger c_{j-1}+c_{j-1}^\dagger c_{j+1})+\sum_j (V_{j+1}-V_j) \alpha_j (c_{j+1}^\dagger c_j+c_j^\dagger c_{j+1}).
\ee
Up to the terms independent of $\mathcal A_\lambda^\ast$ it follows from e.g. Wick's theorem that the action is simply proportional to the a sum of squares of individual contributions in the expression above:
\begin{multline}
\mathcal S(\mathcal A_\lambda^\ast)={\rm const}+{2^L\over 4} \sum_j\left[ \left(\partial_\lambda V_j-2 J(\alpha_j-\alpha_{j-1})\right)^2+2 J^2 (\alpha_{j}-\alpha_{j-1})^2+2 (V_{j+1}-V_j)^2\alpha_j^2\right]\\
={\rm const} +{2^L\over 4}\sum_j \left[ \left(\partial_\lambda V_j \right)^2 +4J \alpha_j \partial_\lambda \left( V_{j+1}-V_j \right) + 6J^2 \left(\alpha_{j}-\alpha_{j-1} \right)^2 +2\left( V_{j+1}-V_j \right)^2 \alpha_j^2  \right]
\end{multline}
Minimizing the action with respect to $\alpha_j$ yields the following set of linear equations
\begin{equation}
-3 J^2\left(\alpha_{j+1}-2\alpha_j +\alpha_{j-1} \right)+\left( V_{j+1}-V_j \right)^2 \alpha_j= -J\partial_\lambda \left( V_{j+1}-V_j \right).
\label{eq:supl_vargauge_discrete}
\end{equation}
This system can be always solved numerically by standard methods. Moreover, whenever the potential is smooth at the level of lattice spacing one can replace discrete differences by continuous derivatives:
\begin{equation}
-3J^2\partial_x^2 \alpha(x)+\left( \partial_x V(x,\lambda) \right)^2 \alpha(x)= -J\partial_\lambda\partial_x V(x,\lambda).
\label{eq:supl_vargauge}
\end{equation}

\subsection{Real CD protocols.}

As in the waiter problem, discussed in detail in the main text, the variational CD term in the Hamiltonian: $\dot\lambda\mathcal A_\lambda^\ast$, can be gauged away by a simple Peierls phase shift (equivalent to the discrete momentum shift). By the way, this is not generally possible for the exact CD driving with the full gauge potential $\dot\lambda \mathcal A_\lambda$. Let us discuss how this can be done explicitly. The local CD Hamiltonian is
\begin{multline}
H_{\rm CD}=H_0+\dot\lambda \mathcal A_\lambda^\ast=-J \sum_j \left[c_{j+1}^\dagger c_j \left(1-i {\alpha_j\dot\lambda\over J}\right)+c_{j}^\dagger c_{j+1} \left(1+i {\alpha_j\dot\lambda\over J}\right)\right]+\sum_j V_j(\lambda) c_j^\dagger c_j\\
=-J\sum_j \sqrt{1+(\dot{\lambda} \alpha_j/J )^2} \left[  e^{-i\phi_j}c_{j+1}^\dagger c_j+e^{i\phi_j}c_{j}^\dagger c_{j+1} \right]+\sum_j V_{j}(\lambda) c_j^\dagger c_j,
\label{eq:h_cd_imag}
\end{multline}
where
\[
\tan \phi_j={\alpha_j\dot\lambda\over J}.
\]
Now we will do a Peierls transformation
\be
c_j\to c_j e^{-i f_j},
\label{eq:phase_rot}
\ee
where $\phi_j=f_{j+1}-f_j$. Under this transformation the CD Hamiltonian clearly becomes real but because the phase is time dependent there is an additional scalar potential term $-\sum_j \dot f_j c_j^\dagger c_j$. Thus the CD Hamiltonian is gauge equivalent to the following real Hamiltonian
\begin{eqnarray}
H_{CD}\sim-\sum_j J_j\left[  c_{j+1}^\dagger c_j+c_{j}^\dagger c_{j+1} \right]+\sum_j U_j c_j^\dagger c_j, 
\label{eq:h_cd_real}
\end{eqnarray}
where the effective potential $U_j$ and the effective hopping $J_j$ are given by
\begin{eqnarray}
J_j&=&J\sqrt{1+(\dot{\lambda} \alpha_j/J )^2}, \\
U_j&=&V_{j,\lambda}-\dot{f_j}=V_{j,\lambda}-\sum_{i=1}^j \frac{J}{J^2+(\dot{\lambda}\alpha_j)^2} \left(\ddot{\lambda} \alpha_j+(\dot{\lambda})^2\partial_\lambda \alpha_j \right).
\end{eqnarray}
As in the waiter problem this protocol only depends on the velocity and acceleration of the potential $\dot\lambda$ and $\ddot\lambda$. Clearly $J_i=J$ and $U_j=V_j$ whenever both the acceleration and the velocity are zero thus at these points both CD Hamiltonians~\eqref{eq:h_cd_imag} and \eqref{eq:h_cd_real} coincide with the lab Hamiltonian. Moreover the phase rotation~\eqref{eq:phase_rot} required to go from the imaginary to the real CD protocol depends only on the velocity $\dot\lambda$ so the wave functions corresponding to these two protocols coincide at zero velocity points. Therefore to implement the real CD protocol~\eqref{eq:h_cd_real} it is preferable to choose a dependence $\lambda(t)$, which has vanishing first and second derivatives in the beginning and in the end of the protocol such that one avoids any discontinuities in the Hamiltonian during the ramp. As in the main text throughout these notes we use
\be
\lambda(t)=\lambda_0+(\lambda_f-\lambda_0)\, \sin^2\left({\pi\over 2} \sin^2\left( {\pi t\over 2\tau}\right)\right),
\label{eq:lambda_t}
\ee
which interpolates between the initial value $\lambda_0$ at time $t=0$ and the final value $\lambda_f$ at time $t=\tau$ satisfying $\dot\lambda(0)=\ddot\lambda(0)=\dot\lambda(\tau)=\ddot\lambda(\tau)=0$.

\subsection*{Linear potential}

\begin{figure}[ht]
  \centering
  \includegraphics[width=0.7\linewidth]{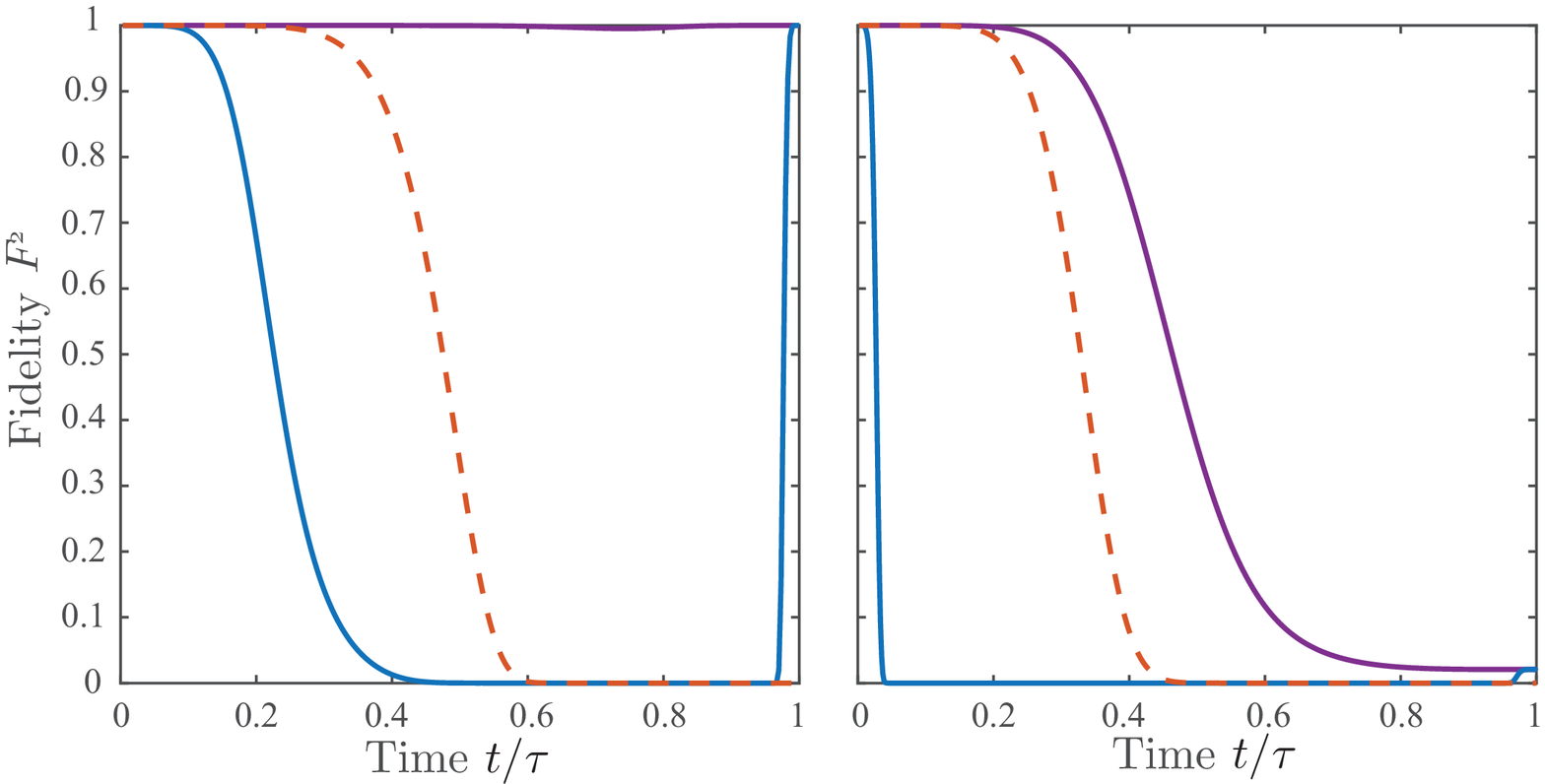}
  \caption{Instantaneous ground state fidelity (squared) $F^2(\tau)$ for two different ramps (left and right) and three different protocols: (i) naive protocol (red dashed line) with $\lambda(t)$ given by Eq.~\eqref{eq:lambda_t}, (ii) imaginary CD protocol (solid purple line) given by Eq.~\eqref{eq:h_cd_imag} and (iii) real CD protocol (blue solid line) given by Eq.~\eqref{eq:h_cd_real}. The system size is $L=512$ at half filling (the total number of fermions is $256$), the total duration of the protocol is $\tau=1/J$ and we fixed $J=1$. Left: the fidelity for the ramp from $\lambda_0=400/L$ and $\lambda_f=40/L$, corresponding to the bulk regime $\kappa L\gg 1$. Perfect fidelity is found for the CD protocol whereas the fidelity of the naive protocol is as low as $3\cdot 10^{-58}$. Right: the fidelity for a protocol that increases the field strength by a factor 10 from $\lambda_0=0.04/L$ to $\lambda_f=0.4/L$ The fidelity for the naive protocol drops to $3\cdot 10^{-16}$ while the CD fidelity decreases only to $2\cdot 10^{-2}$.}
  \label{fig:electric}
\end{figure}

We start from the linear potential $V(x,\lambda)=\lambda x$. In this and the following examples we will assume the potential $V(x,\lambda)$ is smooth on the lattice scale, i.e. $\lambda\gg J$ and one can use the continuum approximation~\eqref{eq:supl_vargauge}. This assumption is not necessary but allows one to simplify the expressions. In all numerical simulations we solve the original discrete equations~\eqref{eq:supl_vargauge_discrete}. For the linear potential, Eq.~\eqref{eq:supl_vargauge} can be solved analytically:
\begin{equation}
\alpha(x)=-\frac{J}{\lambda^2} \left(1 + A \exp\left(-\kappa x \right)+B \exp\left(\kappa x \right)\right),
\end{equation}
where $\kappa=\lambda/(\sqrt{3}J)$ and $A,B$ are arbitrary constants. One can check that for the system with vanishing boundary conditions for fermions the action is minimized by requiring that $\alpha(x)$ also vanishes at the boundary. Physically this corresponds to the absence of the boundary currents in the CD protocol. Then for  $x\in [-L/2, L/2]$ one finds:
\begin{equation}
\alpha(x)=-\frac{J}{\lambda^2} \left(1 - \frac{\cosh(\kappa x)}{\cosh(\kappa L/2)} \right).
\label{eq:gauge_linear}
\end{equation}
For large systems $\kappa L\gg 1$ we see that, except near the boundaries, $\alpha(x)\approx -J/\lambda^2$, which is precisely the exact result we discussed in the main text (q.v.). Near the boundaries the variational solution~\eqref{eq:gauge_linear} is only an approximation and the true gauge potential contains long-range hopping terms. As long as $\kappa L\gg 1$, which physically means that the potential difference across the system $\Delta V=\lambda L$ is much bigger than the hopping $J$, the boundary effects are expected to be unimportant and hence the variational solution should be very good. These results are illustrated in Fig.~\ref{fig:electric}, where we compare  
fidelities squared of naive and CD protocols using the gauge potential~\eqref{eq:gauge_linear} in two different regimes. The left panel corresponds to the ramp of electric field from the initial value $\lambda_0\approx 0.78$ to the final value $\lambda_f\approx 0.078$  (we set $J=1$) for the system of fermions at half filling and the system size $L=512$. The protocol duration is $\tau=1/J$ and we the ramp shape is given by Eq.~\eqref{eq:lambda_t}. In this regime $\kappa L\geq 40/\sqrt{3}$ is large at all times. Then the CD protocol gives nearly unit fidelity (purple line) while the naive protocol gives fidelity $F^2(\tau)\approx 3\times 10^{-58}$, i.e. almost sixty orders of magnitude less. On the right panel we show similar results but for turning on a tiny electric field with $\lambda_0\approx 7.8\cdot 10^{-5}$ and $\lambda_f\approx 7.8\cdot 10^{-4}$. Under these conditions $\kappa L\ll 1$ in the beginning of the protocol and the exact bulk solution is simply irrelevant. The final fidelity of the naive protocol in this regime is $F^2(\tau)\approx 3\cdot 10^{-16}$ while the CD fidelity is $F^2(\tau)\approx 0.02$. Despite the approximate CD protocol is no longer exact even in this maximally unfavorable regime it still gives almost fourteen orders of magnitude gain in the performance compared to the naive protocol. In both panels solid purple (blue) lines show fidelity for the imaginary (real) protocol given by Eqs.~\eqref{eq:h_cd_imag} and  \eqref{eq:h_cd_real} respectively. As we discussed for the real protocol the instantaneous fidelity in the lab frame becomes very low at intermediate times, because the system approximately follows the ground state of a gauge equivalent Hamiltonian. 

If we focus on the regime, where boundaries are not important, i.e. $\kappa L\gg 1$ then the real CD protocol allows for additional simplification. Indeed in this case $\alpha(x)\approx -J/\lambda^2$ is $x$-independent, therefore the renormalization of hopping according to~\eqref{eq:h_cd_real} is also spatially uniform. Because the Hamiltonian can be rescaled in an arbitrary way without affecting transition amplitudes (this rescaling is actually equivalent to rescaling the time units in the moving frame) we can absorb renormalization of hopping into the renormalization of the linear potential such that the real CD protocol becomes equivalent to
\begin{eqnarray}
H_{CD}\sim -J  \sum_j  \left[ c_{j+1}^\dagger c_j+c_{j}^\dagger c_{j+1} \right]+\frac{\lambda}{\sqrt{1+\dot{\mu}^2}}\left(1-\frac{\mu \ddot{\mu}}{1+\dot{\mu}^2} \right)\sum_j c_j^\dagger c_j,  \quad{\rm where}\; \mu=1/\lambda.
\end{eqnarray}
So we see that in this case the (exact up to boundary terms) CD driving amounts simply to modifying the time protocol for turning all the linear potential. In more general situations renormalization of hopping is not uniform and can not be eliminated by any global time transformation.

\subsection*{Inserting and moving Eckart potential}
Let us show some additional results for the setups discussed in the main text. Namely we will analyze inserting and moving the Eckart potential to the gas of free fermions. We will use exactly same parameters as in the main text.

\begin{figure}[h]
  \centering
  \includegraphics[width=0.85\linewidth]{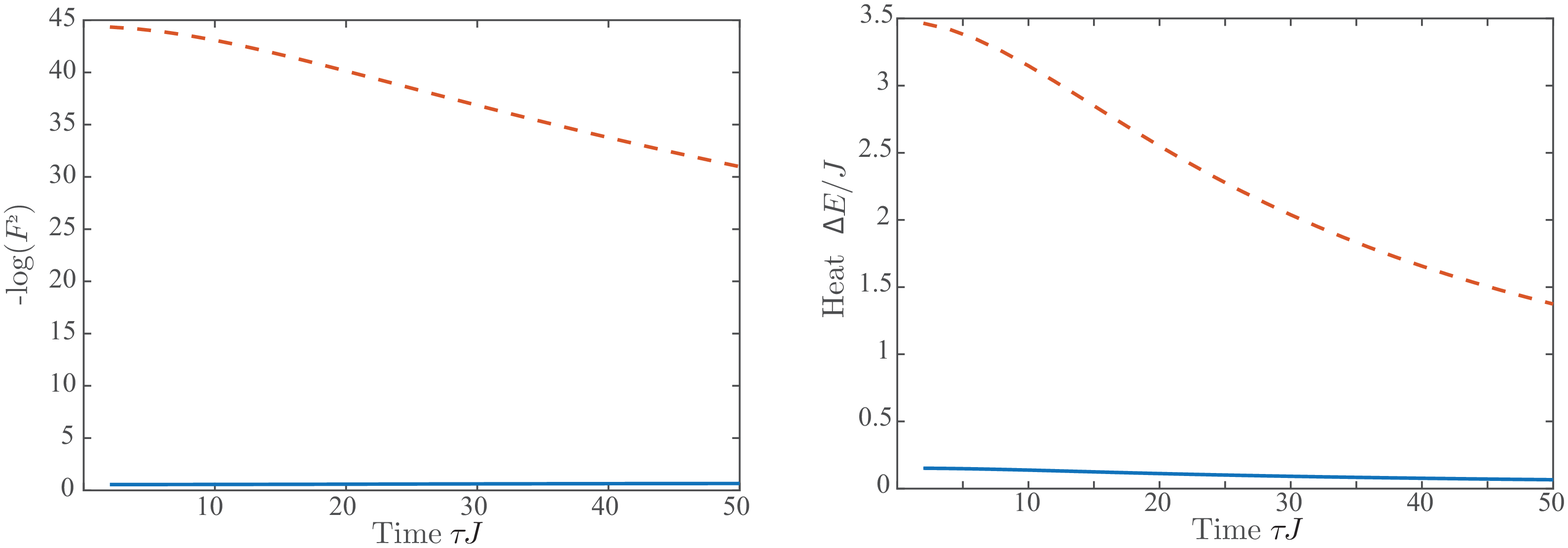}
  \caption{{\bf Scaling inserting potential.} An Eckart potential with $\xi=8$ and final strength $\lambda_f=2J$ is inserted into a fermionic chain of $L=512$, which is half filled. Final (squared) fidelity (left panel) and excess energy (right) panel is shown for the naive and counter-diabatic protocol. }
  \label{fig:inserting}
\end{figure}

Let us consider the protocol where we insert the Eckart potential: 
\be
V(\lambda,j)={\lambda(t)\over \cosh^2 j/\xi},\; j\in[-L/2,L/2],
\ee
where $\lambda(t)$ is given by Eq.~\eqref{eq:lambda_t} with $\lambda_0=0$, $\lambda_f=2J$, $\xi=8$, the system size $L=512$ and the initially we the system is prepared in the ground state of free fermions at half filling. In Fig.~\ref{fig:inserting} we show final fidelity squared $F^2(\tau)$ and the excess energy (heating) $\Delta E$ at the end of the protocol as a function of the ramping time $\tau$. Excess energy is defined as the difference between the actual energy and the instantaneous ground state energy
\[
\Delta E=\langle \psi(\tau)| H_0(\tau) |\psi(\tau)\rangle- E_0^{\rm gs}(\tau).
\]
As we mentioned in the main text, when a potential is inserted into a gas of free fermions, the system suffers from Anderson orthogonality catastrophe. This makes adiabatic loading very difficult: particles should have enough time to rearrange to remove this orthogonality. A naive argument usually exploited in quantum annealing literature for estimating the time required for adiabatic loading is based on the Landau-Zener criterion:
\be
{d\Delta\over dt}\sim \Delta^2,
\ee
where $\Delta$ is the minimum gap in the system. Using that $\Delta \sim J/L$ and estimating $d\Delta /dt\sim \Delta/\tau$ we get a very simple criterion for the adiabaticity:
\be
J \tau\gtrsim L,
\ee
i.e. the loading time for the naive protocol should scale extensively with the system size. At faster speeds one expects very low fidelity close to that of the initial state. This is indeed what we observe numerically. However, CD protocol makes this estimate simply irrelevant as it suppresses transitions between states allowing one to get very high fidelity as shown in the plot. Moreover contrary to the naive protocols this fidelity has very weak dependence on the ramping time. Excess energy shows a very similar behavior as the fidelity (right panel in Fig.~\ref{fig:inserting}). Here the gain is also large, more than an order of magnitude for loading times $\tau\lesssim 10/J$ and remains significant all the way to $\tau=50/J$.

\begin{figure}[h]
  \centering
  \includegraphics[width=0.8\linewidth]{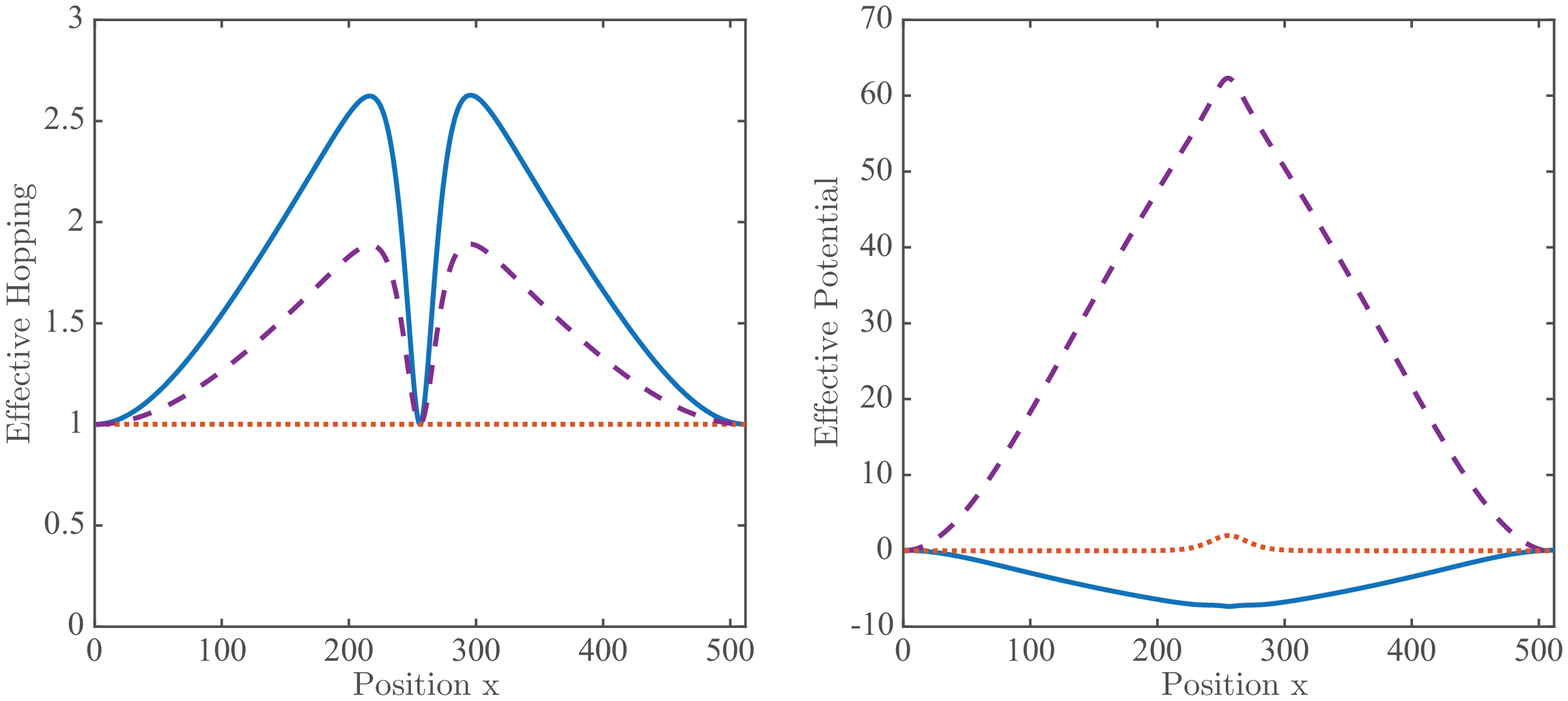}
  \caption{{\bf CD protocol for inserting potential.} When an Eckart potential is inserted in a lattice Fermi gas, renormalization of the hopping and the potential makes the protocol more adiabatic. An effective protocol, $J_i$ and $U_i$, is shown for a protocol lasting $\tau=10/J$ with $\xi=20$ and $\lambda_f=2J$. The purple dashed line is the result at the point of maximum initial acceleration and the blue full line indicates the result at maximum velocity (middle of the protocol). The left panel shows the effective hopping. One should think of this as a sort of refractive index for the fermions, slowing down scattered particles. Note that this term is only sensitive to the speed of the protocol and not to acceleration. The right panel shows the effective potential. In the middle of the protocol (blue) line, the acceleration vanishes so the entire correction to the naive potential is due to the velocity term. Note that, similar to the electric field protocol, one has to drive in the opposite direction for a while.   The red dotted lines show the naive hopping and potential halfway down the protocol.}
  \label{fig:insertingprotocol}
\end{figure}
\subsubsection*{Moving a scattering potential}

\begin{figure}[ht]
  \centering
  \includegraphics[width=\linewidth]{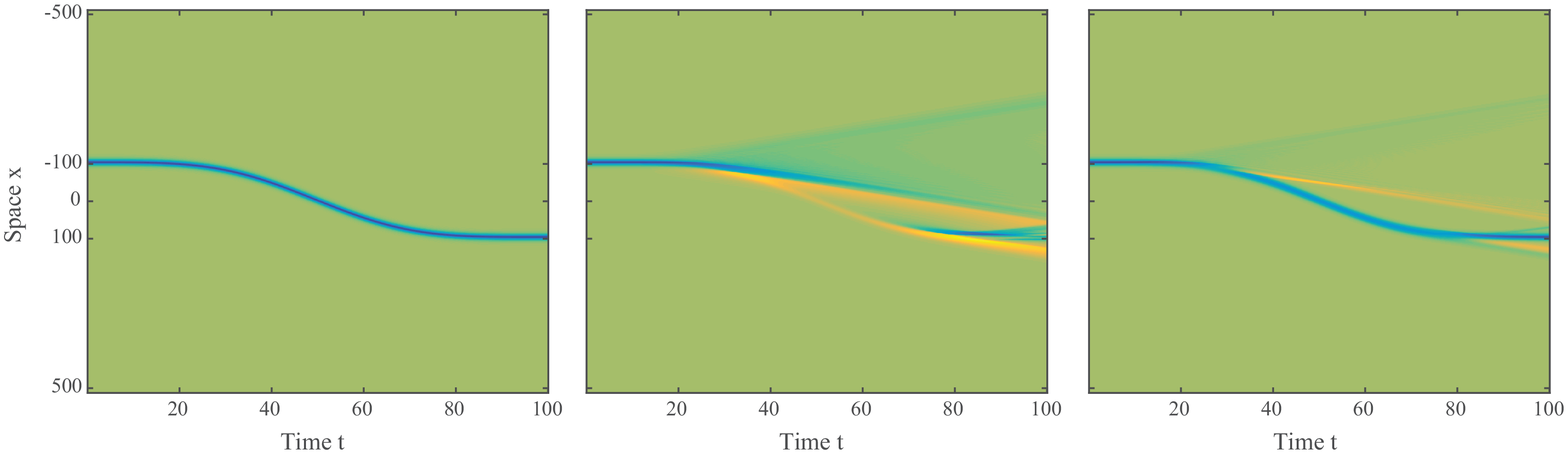}
  \caption{{\bf Moving an obstacle.} A scatterer, modeled by an Eckart potetial with $V_0=2$ and $\xi=8$, is displaced through a half-filled fermi sea. An adiabatic protocol would display the density as depicted in the left most figure, where the dip in the density simply moves according to protocol \eqref{eq:lambda_t}. The actual naive protocol, shown in the middle figure, however generates a large amount of particle-hole excitations, in particular at the points where the protocol accelerates. By local counter-diabatic driving most of these excitations can actually be removed as shown in the right most figure.  }
  \label{fig:densityimpurity}
\end{figure}

The same Eckart potential can also be moved through the sample:
\[
V(\lambda,j)={V_0\over  \cosh^2 (j-\lambda)/\xi},
\]
where $\lambda$ now stands for for the position of the potential maximum. We fix $V_0=2J$, $\xi=8$ and the system size $L=1024$. In Fig.~\ref{fig:densityimpurity} we show electron density as a function of time for a protocol where the center of the potential $\lambda$ moves from the initial value $\lambda_0=-100$ to the final value $\lambda_f=100$. The right panel shows the instantaneous ground state density, which simply tracks the position of the potential. The second panel shows the density of the naive protocol. We can clearly see the excess density in front of the potential and the depleted density region behind the potential as expected. For the parameters shown the final fidelity of the naive protocol is essentially zero, i.e. $F^2(\tau)\approx 5\cdot 10^{-128}$. The right panel shows the fermion density for the CD protocol given by~\eqref{eq:H_CD_1Dfermions}. This protocol visibly shows much fewer excitations and consequently much smaller energy dissipation and much higher fidelity $F^2(\tau)\approx 4\cdot 10^{-5}$. As in the previous example achieving so high fidelity is simply unthinkable for such large system sizes and such fast rates.

\begin{figure}[h]
  \centering
  \includegraphics[width=0.8\linewidth]{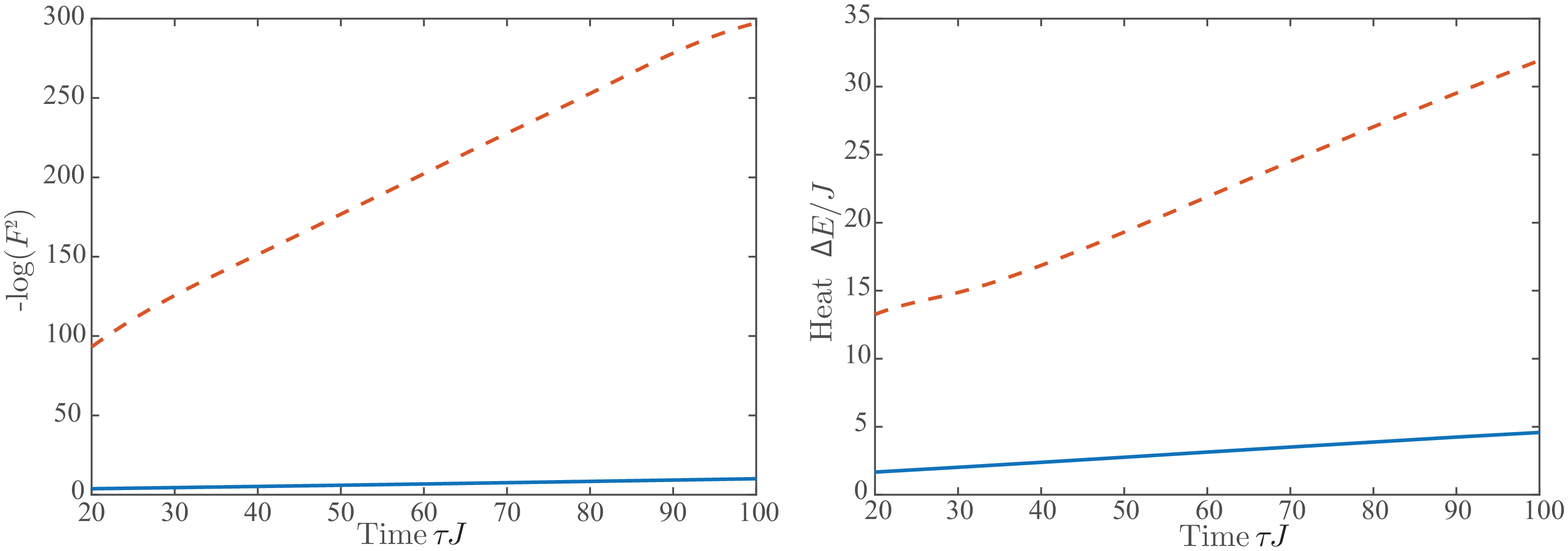}
  \caption{{\bf Scaling moving potential.} An Eckart potential with $\xi=8$ and with a strength of $V_0=2$ is dragged through a fermionic chain of $L=1024$, which is half filled. Average speed $v=\Delta L/\tau$ is kept fixed. Final (squared) fidelity (left panel) and excess energy (right) panel is shown for the naive and counter-diabatic protocol. }
  \label{fig:moving}
\end{figure}

In Fig.~\ref{fig:moving} we show the fidelity (left) and the excess energy (right) as a function of the duration of the protocol. While extending the duration, the average speed at which the potential is moved is kept fixed. We see a linear increase both in excess energy and logaritmic fidelity, consisted with standard friction force. Note that at zero temperature there is no linear (viscous) friction proportional to the velocity but there is always a nonlinear friction force (pressure drag) scaling as $\dot\lambda^2$ , which physically comes from the moving potential scattering fermions. 
As expected, both the naive and CD protocols show increasing fidelity and heating as a function of time $\tau$. However, again the CD driving gives dramatic improvements over the naive protocol.

\begin{figure}[h]
  \centering
  \includegraphics[width=0.95\linewidth]{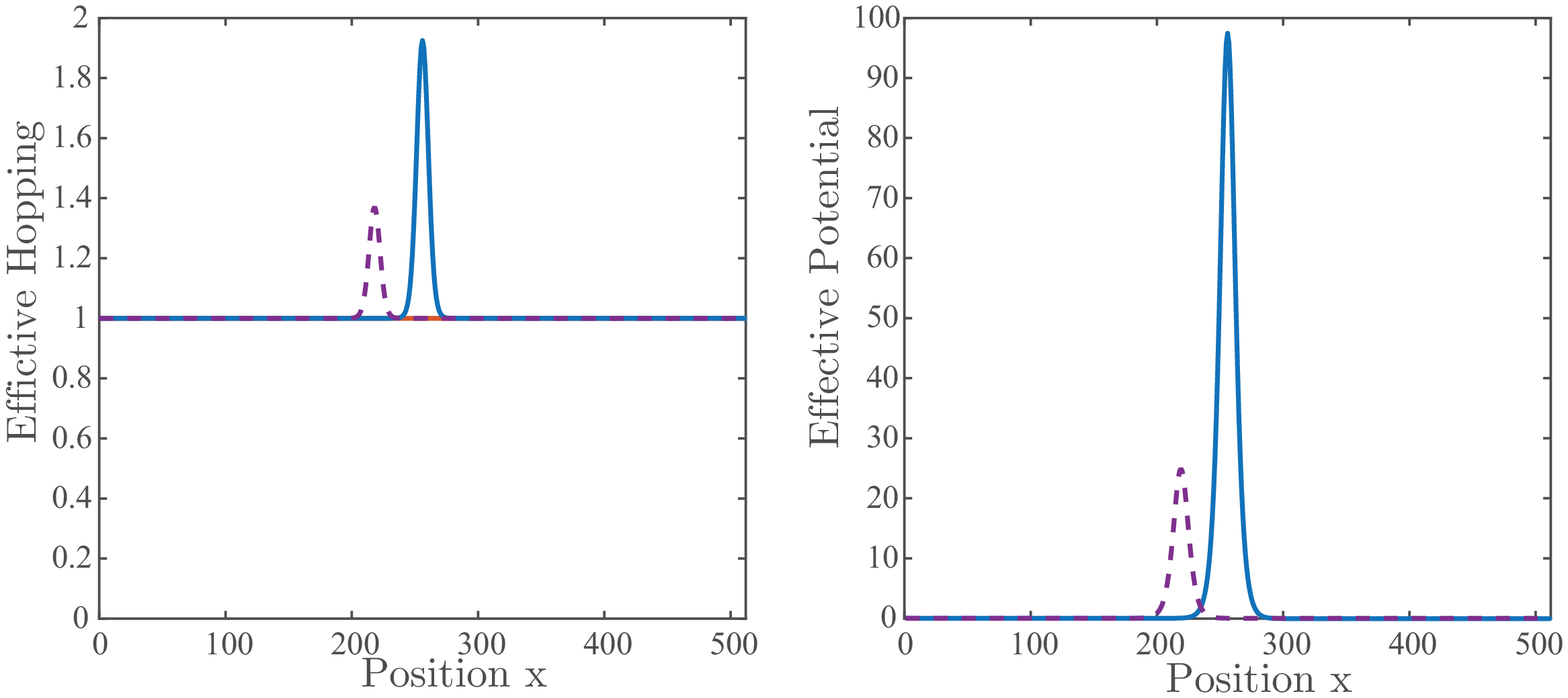}
  \caption{{\bf CD protocol moving potential.}  One can significantly reduce the friction on an scattering potential that is dragged through a Fermi-gas by renormalizing the hopping and the shape of the scattering potential. The effective hopping and potential, $J_i$ and $U_i$, are shown for an Eckart potential with $V_0=2J$, $\xi=8$, $\tau=100/J$ and $\Delta X=100$. The blue lines denote the result halfway down the protocol, where the velocity is maximal and the dashed purple line shows the result at the initial point of maximal acceleration. At this speed, on average $v=1J$, both the hopping and the effective potential are dominated by this term. Closer investigation however reveals that the effective potential, unlike the hopping, is asymmetric due to the acceleration.}
  \label{fig:movingprotocol}
\end{figure}

\section{Optimal local counter-diabatic gauge for ergodic spin chain}
Let's consider a uniformly driven quantum spin chain with the Hamiltonian
\begin{equation}
H_0=\sum_j \left( J(\lambda(t)) \sigma_j^z\sigma_{j+1}^z +Z_j(\lambda(t)) \sigma_j^z +X_j(\lambda(t)) \sigma_j^x \right),
\label{eq:H_spin}
\end{equation}
where $\lambda(t)$ specifies some path in the coupling space. Note that the Hamiltonian can be always rescaled by an arbitrary factor so there are only two independent couplings. The adiabatic gauge potential $\mathcal A_\lambda$ should be Hermitian and imaginary, so has to contain odd number of Pauli matrices $\sigma_j^y$. The most local variational gauge potential $\mathcal A_\lambda^\ast$ is thus of the following form
\begin{equation}
\mathcal A_\lambda^\ast= \sum_j \alpha_j \sigma_j^y.
\end{equation}
In the next order of approximation one can add terms like $\sigma_j^y \sigma_{j+1}^{x,z}$ and so on. It is easy to see that
\be
 G_\lambda\equiv \partial_{\lambda} H+i [\mathcal A_\lambda^\ast, H]=\sum_j \left( (X_j'-2 Z_j \alpha_j) \sigma^x_j+ 2\alpha J (\sigma^x_j \sigma^z_{j+1}+\sigma^z_j \sigma^x_{j+1})+(Z_j'+2X_j \alpha_j) \sigma^z_j+J'  \sigma_j^z\sigma_{j+1}^z \right),
\ee
where `prime' stands for the derivative with respect to $\lambda$. As with the free fermion case computing the Hilbert-Schmidt norm of this operator is trivial and essentially amounts to adding up squares of coefficients in front of independent spin terms:
\be
\tr(G_\lambda^2)=2^L \sum_j \left((X_j'-2 Z_j \alpha_j)^2+8\alpha_j^2 J^2+(Z_j'+2X_j \alpha_j)^2+(J')^2 \right).
\ee
Minimizing this with respect to $\alpha$ we find
\be
\alpha_j={1\over 2}{Z_j X_j'-X_j Z_j'\over Z_j^2+X_j^2+2J^2}.
\ee
 For $J=0$ this gauge potential is exact, as it is simply a generator of spin-rotations in $x-z$ plane. However for finite $J$ this potential is only approximate.
 
\begin{figure}
  \centering
  \includegraphics[width=0.5\linewidth]{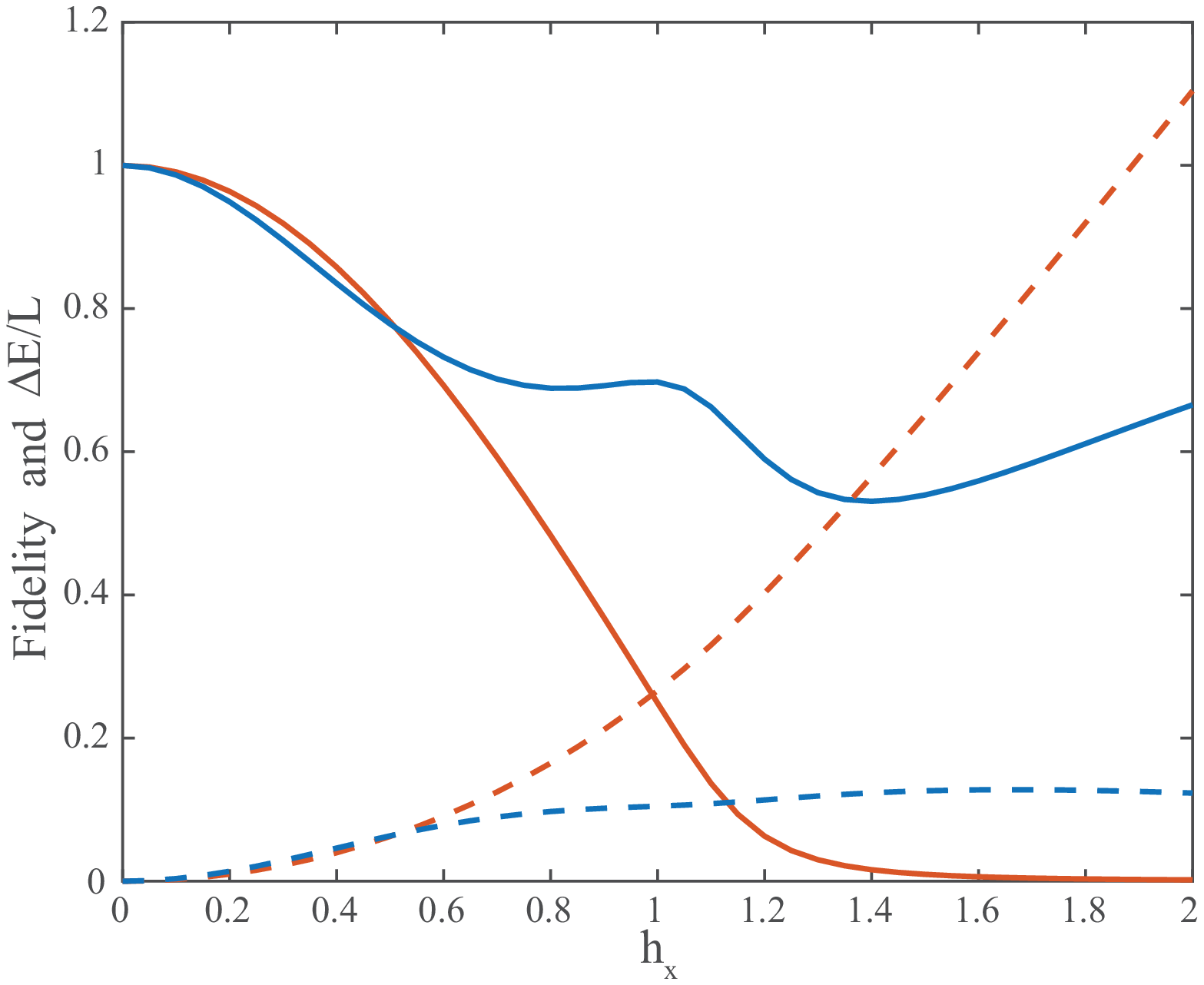}
  \caption{{\bf Spin chain characteristics.} A chain of 15 spins driven from a product state of all up spins to the ground state of Hamiltonian \eqref{eq:H_spin}. In particular, we start from $J=h_x=0$ and $h_z=0.02$ and ramp $J$ to -1 and $h_x$ to 2, using protocol \eqref{eq:lambda_t}, while keeping $h_z$ fixed.  The squared fidelity (full lines) and excess energy density (dashed lines) are shown for the naive and CD protocols, in red and blue respectively.}
  \label{fig:spinchain}
\end{figure}

To complement the discussion in the main text let us analyze an annealing protocol with the goal to try to prepare the spins in the ground state of the Hamiltonian \eqref{eq:H_spin} out of the initial product state of all up spins by driving both the spin coupling $J$ and the $x$-magnetic field from zero to a finite value while keeping $h_z$ fixed. The squared fidelity for such a protocol for a chain consisting of 15 spins is shown in Fig.~\ref{fig:spinchain}. In this plot we fix the final value of $J=-1$ and vary the final value of $h_x$. Both $J$ and $h_x$ are increased from zero according to the protocol~\eqref{eq:lambda_t}. The field $h_z=0.02$ is kept at small constant value. Note that in thermodynamic limit and at $h_z=0$ this system undergoes a quantum phase transition at $h_x=J$ so we test the CD driving protocol in a proximity to the quantum critical point, where non-adiabatic effects are enhanced due to the Kibble-Zurek mechanism. For small values of $h_x<1$ the final ground state is ferromagnetic. Since we started from an all up state, we reach the final state with high fidelity even without CD driving. Moreover, in this case the CD protocol is actually slightly worse than the naive protocol, presumably because it does not target specifically the ground state of the system. For large $h_x\gtrsim 1$ the naive protocol does a very poor job in converting the all up state into the paramagnetic $x$ state. Note that deep in the paramagnetic regime the coupling between the spins is irrelevant and so the CD protcol performs great because it is exact in the Landau-Zener limit ($J\to 0$).  Remarkably, the CD protocol remains efficient in preparing critical states close to $h_x=1$. Let us mention that, while the final point in the protocol is close to integrable $h_z=0$, at intermediate times the system is very far from integrability. 

\begin{figure}[t]
  \centering
  \includegraphics[width=0.9\linewidth]{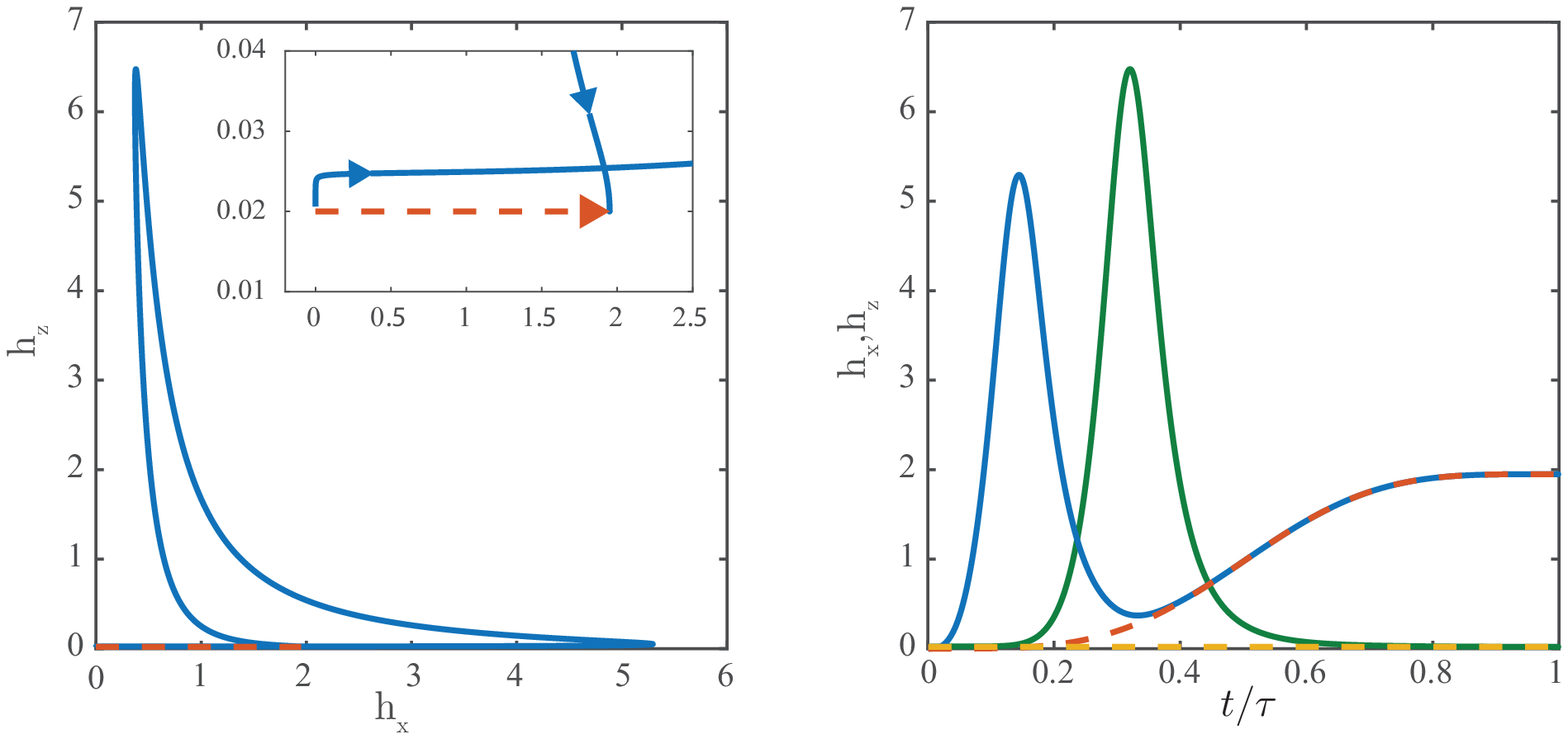}
  \caption{{\bf Annealing protocol for a spin spin chain.} An example of a particular annealing protocol that turns a tensor product state of all up spins into the ground state of  \eqref{eq:H_spin}. We start from $J=h_x=0$ and $h_z=0.02$ and ramp $J$ to 1 and $h_x$ to 2 according to Eq.~\eqref{eq:lambda_t} while keeping $h_z$ fixed. The time-dependence of the magnetic fields is depicted in the right panel and the left panel shows the trajectory in the $h_x-h_z$-plane. The time dependence of $J$ is not shown, as it is the same in the naive and the CD protocol. \emph{Panel A:}The dashed red line is the naive protcol, that goes in a straight line from the initial to final point. The counter-diabatic protcol, shown in full blue, takes a large detour. \emph{Panel B:} Dashed red line shows the dependence of the naive $h_x$ on time, the full blue line is the counter-diabatic $h_x$. Yellow dashed line depicts a constant $h_z$ in the naive protcol. In contrast, the couter-diabatic protcol, shown in green, shows a large peak. For a chain of 15 spins, this counter diabatic protocol has a probability of $0.66$ of ending in the ground state whereas the naive protcol only ends in the ground state with a probability $2.1\, 10^{-3}$. }
  \label{fig:spinprotocol}
\end{figure}

\begin{figure}
  \centering
  \includegraphics[width=0.5\linewidth]{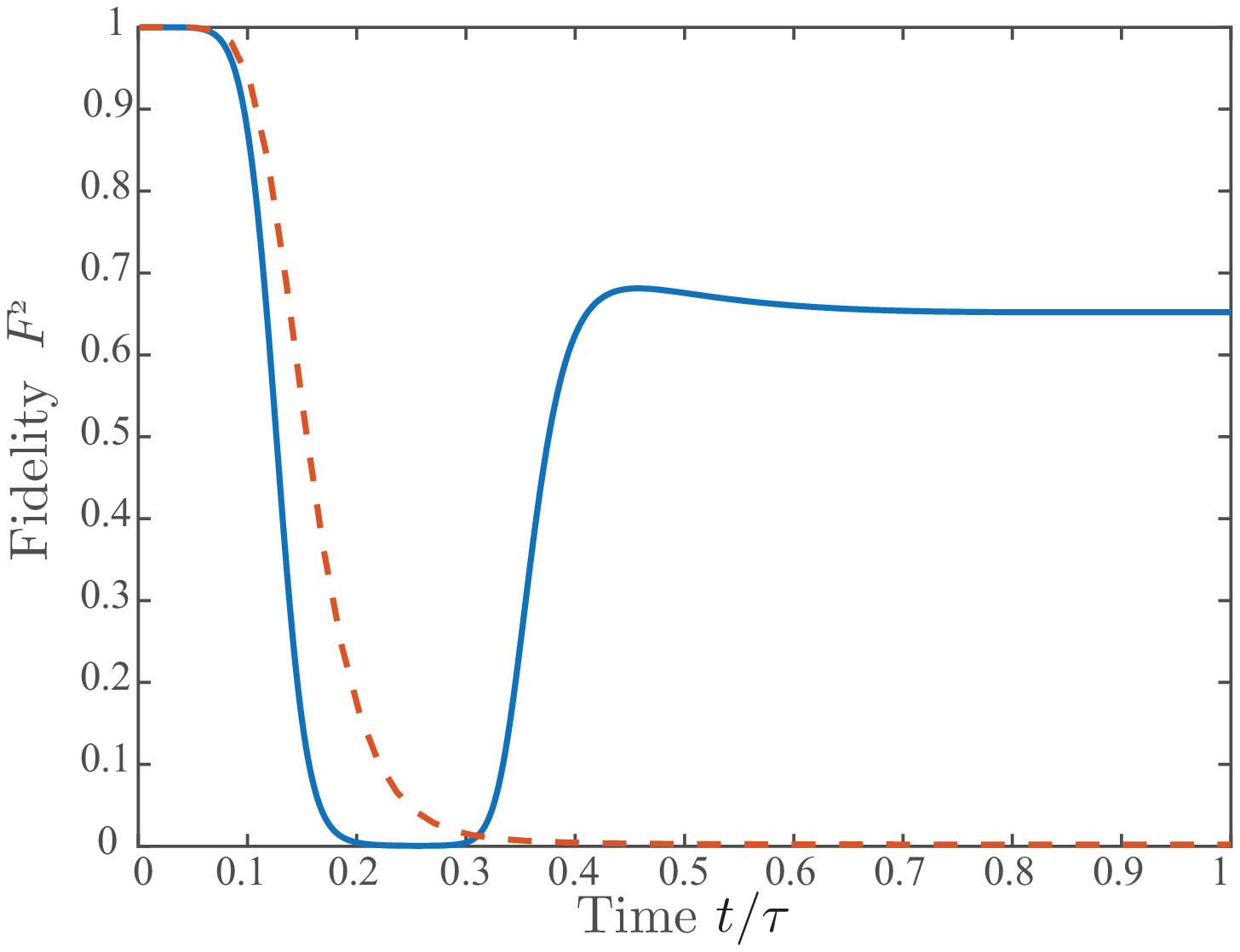}
  \caption{{\bf Instantaneous fidelity spin chain.} During the annealing protocol from Fig.~\ref{fig:spinprotocol} the probability to be in the instantaneous ground state changes with time. For the naive protocol, shown in dashed red, this probability rapidly drops to a small value (of about $2.1\, 10^{-3}$). The counter-diabatic protocol shows an ever faster drop but quickly revives and ends up close to the ground state. Note that this drop is of linked to the peak in the x-magnetic field in the protocol (see Fig.~\ref{fig:spinprotocol}).}
  \label{fig:spinfidt}
\end{figure}

For completeness, let us note that one can eliminate the $\sigma^y$ term by rotating everything around the z-axis. This is similar to the gauge transformation for the fermions. Indeed, the actual CD Hamiltonian is given by
by\begin{equation}
H_{CD}=\sum_j \left( J \sigma_j^z\sigma_{j+1}^z +Z_j \sigma_j^z +X_j \sigma_j^x+Y_i\sigma_j^y \right)  \quad{\rm with}\; Y_i=\dot{\lambda}\alpha_i.
\end{equation}
By applying the unitary $U=\exp(i\theta\sigma^z/2)$ over the right angle $\tan \theta=Y/X$ elimates the y-field from the Hamiltonian. Since the angle is time-dependent, an additional field $\dot{\theta}/2\sigma^z$ is introduced for every spin, resulting in 
\begin{equation}
H_{CD}\sim \sum_j \left( J \sigma_j^z\sigma_{j+1}^z +\left(Z_j+\frac{1}{2}\frac{X_j\dot{Y}_j-Y_j\dot{X}_j}{X_j^2+Y^2_j}\right) \sigma_j^z +\sqrt{X_j^2+Y_j^2} \sigma_j^x\right) 
\end{equation}
Remarkably this Hamiltonian is structurally equivalent to the original one. This means that, similar to the electric-field example for the fermions, one can significantly increase the fidelity to reach the final target state by simply using a different protocol than the naive one. An example of such protocol is shown in Fig.~(\ref{fig:spinprotocol}) and Fig.~(\ref{fig:spinfidt}) . As before, this protocol gives very low fidelity at intermediate times because the system approximately follows the ground state of a rotated Hamiltonian.

\end{document}